\documentclass[review]{elsarticle}
\makeatletter
\def\ps@pprintTitle{%
 \let\@oddhead\@empty
 \let\@evenhead\@empty
 \def\@oddfoot{\centerline{\thepage}}%
 \let\@evenfoot\@oddfoot}
\makeatother

\biboptions{numbers,sort&compress}
\pdfoutput=1
\usepackage{graphicx}
\usepackage{amsfonts, amsmath, amssymb}
\usepackage{mathabx}
\usepackage{booktabs,siunitx}
\usepackage[svgnames,table]{xcolor}
\usepackage[tableposition=above]{caption}
\usepackage{pifont}
\usepackage{bm}
\usepackage{cancel}
\usepackage{caption}
\usepackage{color}
\usepackage{enumerate}
\usepackage{float}
\usepackage{hyperref}
\usepackage{soul}
\usepackage[english]{babel}
\usepackage[margin=1in]{geometry}
\usepackage{mathtools}
\usepackage{multirow}
\usepackage{setspace}
\usepackage{subfigure}
\usepackage{stmaryrd}
\usepackage{lineno}
\usepackage{tabularx}
\DeclareUnicodeCharacter{00B3}{\textsuperscript{3}}
\usepackage[ruled,linesnumbered]{algorithm2e}
\RestyleAlgo{ruled}
\SetKwComment{Comment}{/* }{ */}

\SetCommentSty{mycommfont}
\DeclareMathAlphabet{\mathpzc}{OT1}{pzc}{m}{it}

\renewcommand \d[1]{{\rm{d}} #1}
\newcommand \D [2]{\frac{\partial #1}{\partial #2}}

\renewcommand{\vec}[1]{\bm{\mathrm{#1}}}
\newcommand{\V}[1]{\bm{\mathrm{#1}}}

\newcommand{\bsigma}{\boldsymbol{\sigma}}

\newcommand{\etal}{\emph{et al. }}

\newcommand{\ncells}{n_{\textrm{cells}}}
\newcommand{\comsol}{COMSOL Multiphysics\textsuperscript{\textregistered}}
\newcommand{\ansys}{ANSYS\textsuperscript{\textregistered}}

\newcommand{\rhozero}{\rho_0}
\newcommand{\rhoone}{\rho_1}

\newcommand{\sigmatwo}{\bm{\sigma}_2}

\newcommand{\pone}{p_1}

\newcommand{\frad}{\V{F}^\textrm{rad}}

\newcommand{\volume}{\makebox[1pt][l]{$-$}V}

\def \div{\nabla \cdot \mbox{}}

\def \bnabla{\V{\nabla}}

\def \x{\vec{x}}
\def \e{\vec{e}}

\def \n{\vec{n}}
\def \r{\vec{r}}

\def \v{\vec{v}}
\def \vone{\vec{v}_1}
\def \pone{p_1}
\def \vpone{\vec{p}_1}
\def \vptwo{\vec{p}_2}

\def \rhoone{\rho_1}
\def \vonehat{\widehat{\vec{v}}_1}
\def \ponehat{\widehat{p}_1}

\def \vtwo{\vec{v}_2}

\def \e{\vec{e}}
\def \iota{\imath}

\def \F{\vec{F}}

\def \U{\vec{U}}

\def \A{\vec{A}}
\def \b{\vec{b}}

\def \rv{\vec{r_v}}
\def \rp{\vec{r_p}}

\def \F{\vec{F}}

\def \G{\vec{G}}

\def \Lmu{\vec{L_{\mu}}}
\def \Drho{\vec{D_{\rho}}}
\def \Llamb{\vec{L_{\lambda}}}
\def \vrho{\vec{\rho}}

\def \n{\vec{n}}
\def \v{\vec{v}}
\def \vvarphi{\vec{\varphi}}

\def \x{\vec{x}}

\def \ev{\vec{e_v}}
\def \ep{\vec{e_p}}

\def \div{\nabla \cdot \mbox{}}

\def \dx{\Delta x}
\def \dy{\Delta y}

\def \Omegaf{\Omega_{\rm f}}
\def \Omegab{\Omega_{\rm b}}

\def \presdisp{\vec{d}_1}
\def \pfunc{p}
\def \half{\frac{1}{2}}
\def \3half{\frac{3}{2}}

\def \etal{\emph{et al.}}
\def \voneb{\vec{v}_{\rm 1b}}
\def \vtwob{\vec{v}_{\rm 2b}}
\def \vb{\vec{v}_{\rm b}}
\def \fb{\vec{f}_\textrm{b}}
\def \fbone{\vec{f}_\textrm{1b}}
\def \fbtwo{\vec{f}_\textrm{2b}}

\usepackage{acronym}
\acrodef{ARF}[ARF]{acoustic radiation force}

\newcommand{\upperRomannumeral}[1]{\uppercase\expandafter{\romannumeral#1}}

\newcommand{\Revone}[1]{{\color{black}#1}}
\newcommand{\Revtwo}[1]{{\color{black}#1}}

\begin{document}
\let\today\relax
\let\underbrace\LaTeXunderbrace
\let\overbrace\LaTeXoverbrace

\begin{frontmatter}

\title{Simulating acoustically-actuated flows in complex microchannels using the volume penalization technique}




\author[UNL]{Khemraj Gautam Kshetri}
\author[SDSU]{Amneet Pal Singh Bhalla\corref{mycorrespondingauthor}}
\ead{asbhalla@sdsu.edu}
\author[UNL]{Nitesh Nama\corref{mycorrespondingauthor}}
\ead{nitesh.nama@unl.edu}

\address[UNL]{Department of Mechanical \& Materials Engineering, University of Nebraska-Lincoln, Lincoln, NE}
\address[SDSU]{Department of Mechanical Engineering, San Diego State University, San Diego, CA}
\cortext[mycorrespondingauthor]{Corresponding author}

\begin{abstract}
We present a volume penalization technique for simulating acoustically-actuated flows in geometrically complex microchannels. Using a perturbation approach, the nonlinear response of an acoustically-actuated compressible Newtonian fluid moving over obstacles or flowing in a geometrically complex domain is segregated into two sub-problems: a harmonic first-order problem and a time-averaged second-order problem, where the latter utilizes forcing terms and boundary conditions arising from the first-order solution. This segregation results in two distinct volume penalized systems of equations. The no-slip boundary condition at the fluid-solid interface is enforced by prescribing a zero structure velocity for the first-order problem, while spatially varying Stokes drift---which depends on the gradient of the first-order solution---is prescribed as the structure velocity for the second-order problem. The harmonic first-order system is solved via MUMPS direct solver, whereas the steady state second-order system is solved iteratively using a novel projection method-based preconditioner. The preconditioned iterative solver for the second-order system is demonstrated to be highly effective and scalable with respect to increasing penalty force and grid resolution, respectively. A novel contour integration technique to evaluate the acoustic radiation force on an immersed object is also proposed. This technique circumvents the use of velocity derivatives within the smeared region. The contour integral is specifically tailored to Cartesian grids. Through a series of test cases featuring representative microfluidic geometries, we demonstrate excellent agreement between the volume penalized and body-fitted grid solutions for the primary first- and second-order fields as well as for the acoustic radiation force that depends on the gradients of these fields. We also identify suitable penalty factors and interfacial smearing widths to accurately capture the first- and second-order solutions. \Revone{These results provide empirical evidence of the efficacy of the volume penalization method for simulating acoustic streaming problems that have commonly been analyzed using body-fitted methods in the acoustofluidic literature.}
\end{abstract}



\begin{keyword}
\emph{fluid-structure interaction} \sep \emph{perturbation technique} \sep \emph{acoustic streaming}
\sep \emph{projection method}



\end{keyword}

\end{frontmatter}




\section{Introduction}
Computational fluid dynamics simulations have become essential for enhancing our understanding of physical phenomena involving the interactions of viscous fluid flows with immersed objects. For simulating complex objects within fluid flows, two classes of numerical methods have emerged: body-fitted or conformal mesh methods and non-conformal mesh methods such as the immersed boundary (IB) method. The former class of methods relies on conformal discretization of fluid and solid domains with mesh elements aligned at the interface. This allows easy boundary condition enforcement at the fluid-solid interface. In contrast, IB methods embed the immersed object within larger, regular computational domains. They enforce boundary conditions via the introduction of appropriate volumetric forcing terms in the governing equations. 

The volume penalization (VP) method, also referred to as Brinkman penalization technique, is a specific type of IB method that treats the embedded solid domain as a porous media with an extremely low permeability. Here, the influence of solid boundaries on fluid flow is modeled by adding a penalization term to the governing equation, mimicking the Brinkman term from porous media theory. Specifically, velocity continuity at the fluid-solid interface is imposed via a volumetric penalty force that is inversely proportional to the body's permeability $\kappa$. Given the use of a single, regular Eulerian grid, VP methods offer several inherent advantages in terms of their simplicity, ease of parallelization, and ability to handle complex boundaries. Since being first proposed by Arquis and Caltagirone~\cite{arquis1984conditions}, these methods have been significantly extended and refined. For example, to achieve high-order (up to fourth-order) spatial accuracy of the penalized solution, Kou et al.~\cite{kou2022immersed} used a combination of VP and high-order flux reconstruction techniques in a discontinuous Galerkin framework for simulating compressible flows. Phase change problems have also been modeled using the VP method, which is typically referred to as Carman-Kozeny drag models in the literature~\cite{carman1997fluid,voller1987fixed}. In melting and solidification problems, a volumetric penalty/drag force is applied to retard solid phase motion~\cite{voller1987fixed,huang2022consistent,thirumalaisamy2023lowmach,thirumalaisamy2025consistent}.

In the literature, the VP method is most commonly used to simulate fluid-structure interaction (FSI). This technique has been extensively applied to modeling FSI in single-phase (uniform density and viscosity) incompressible flows~\cite{bergmann2011modeling,engels2015numerical,gazzola2011simulations}. Recently, VP has also become popular for analyzing multiphase FSI, such as solid motion in incompressible gas-liquid flows. Some examples include water entry/exit problems for estimating hydrodynamic loads for solid bodies slamming into air-water interfaces~\cite{bhalla2020entryexit}, FSI and optimal control of wave energy converters~\cite{khedkar2021iswec,khedkar2022model,khedkar2024preventing}, and modeling the hydrodynamics of a dolphin jumping out of water~\cite{bergmann2022numerical}. 

In this article, we explore the use of the VP method as a promising approach for simulating acoustofluidic phenomena. Acoustofluidics, which refers to the merger of microfluidics and acoustics, has emerged as a promising technology for a range of exciting applications, including biomedical diagnostics, lab-on-a-chip systems, tissue engineering, and point-of-care devices~\cite{bose2015role,ding2013surface,laurell2014microscale,rasouli2023acoustofluidics}. Acoustofluidic systems are typically characterized by the propagation of high-frequency acoustic waves through fluids. The high-frequency acoustic interaction with viscous fluids not only induces a corresponding high-frequency response of the fluid but also results in nonlinear flow rectification that leads to time-averaged \emph{streaming flow} and \emph{acoustic radiation force} on immersed objects. Combined, the streaming flow and acoustic radiation force have been used for precise, non-contact manipulation of fluids and particles to enable important functionalities at microscales such as fluid mixing~\cite{huang2015acoustofluidic,huang2013acoustofluidic,ozcelik2014acoustofluidic}, pumping~\cite{junahuang2014reliable,pavlic2023sharp}, chemical gradient generation~\cite{ahmed2013tunable,huang2018sharp,huang2015spatiotemporally}, particle sorting~\cite{nawaz2015acoustofluidic}, patterning~\cite{chen2013tunable,guo2015controlling}, and microscale locomotion~\cite{dillinger2024steerable,dillinger2021ultrasound,ren20193d,mcneill2020wafer}. To complement the experimental advances in acoustofluidics with enhanced physical understanding, several studies concerning the physical mechanism and mathematical modeling of these systems have been reported~\cite{kshetri2024evaluating,pavlic2022influence,muller2013ultrasound,barnkob2018acoustically,baasch2018acoustofluidic}. The acoustofluidic problem is typically formulated via a perturbation approach, which expands the flow variables in terms of a small Mach number. This results in two linear sub-problems: a time-harmonic first-order system governing the oscillatory response, and a steady second-order system governing the time-averaged response. A vast majority of the existing numerical studies~\cite{muller2012numerical,nama2014investigation,pavlic2022influence,ovchinnikov2014acoustic,chen2018three} on acoustofluidic systems are two dimensional and rely on commercial solvers (e.g., \comsol, \ansys), with limited attention given to the scalability of these approaches for large problems. Additionally, these models typically employ body-fitted grids to study streaming in microchannels. Given the proliferation of complex geometrical features within experimental systems, IB methods---which enable the use of structured grids and facilitate efficient parallelization---offer a promising alternative for acoustofluidic simulations. \Revone{However, to the best of our knowledge, the use of immersed boundary (IB) or fictitious domain methods for modeling acoustofluidic systems remains limited. Previous studies employing IB methods for acoustically actuated flows have primarily focused on modeling the acoustic wave response through direct numerical simulations of compressible flow, which differs from the perturbation-based approach for compressible flows adopted in this work~\cite{wang2020immersed,seo2011high,bilbao2022immersed,bilbao2023modeling,lemke2023approximate}. In contrast, Sustiel and Grier~\cite{sustiel2024dynamics} proposed a perturbation-based IB approach to study forces on an acoustically-levitated fluid droplet. However, their formulation does not account for the influence of channel geometry and employs a hybrid IB approach that uses a semi-analytical solution of the acoustic scattering problem.} \Revtwo{More recently, Xie~\etal~\cite{xie2025numerical} reported an immersed boundary approach to compute acoustic radiation force, albeit without considering the second-order velocity field.} Recently, we have reported a perturbation-based, variable-coefficient acoustofluidic solver to study streaming flows and verified its order of accuracy via manufactured solutions~\cite{kshetri2025consistent}. As part of our ongoing efforts to develop a large-scale parallel, multiphase/multicomponent acoustofluidic solver capable of handling immersed bodies, the current work incorporates the VP method to model obstacles and geometrical features within acoustically actuated microfluidic flows. 

An important consideration in applying the VP method to acoustofluidic problems is its compatibility with the perturbation approach. Specifically, the introduction of a penalty term into the compressible Navier–Stokes equations via the VP method and the subsequent perturbation expansion results in two distinct penalty terms, one for each order of the system. The time-harmonic first-order system imposes a uniform, homogeneous Dirichlet boundary condition at the fluid-solid interface, whereas the steady state second-order system requires a spatially-varying velocity boundary condition derived from the numerically obtained, first-order velocity solution. Given the diffuse interface representation of the fluid-solid interface in the VP method, this method is only first-order accurate with respect to spatial discretization. As such, the accuracy of the VP method for the second-order problem that requires the imposition of derivatives of the first-order velocity field remains unclear and needs to be assessed for both smooth and non-smooth problems. Further, while VP methods have been extensively employed for time-dependent simulations, their application to steady-state problems remains relatively underexplored in the literature. Recently, Thirumalaisamy et al.~\cite{thirumalaisamy2021critique} employed a VP technique to solve the steady-state Poisson equation on complex domains with spatially varying Neumann and Robin boundary conditions. Aguayo and Lincopi~\cite{aguayo2022analysis} solved steady-state Stokes problem using the VP method and provided convergence analysis and error estimates for this method. In order to simulate acoustofluidic flows in complex microchannels, the VP method needs to be applied to both steady-state and time-harmonic systems.

In this work, we employ the VP method for the time-harmonic first-order and steady-state second-order system of equations for simulating acoustic streaming in complex microchannels.
Our results provide empirical evidence that the VP method provides excellent accuracy even for the second-order streaming problem. This is confirmed by comparing VP solutions to body-fitted grid solutions. We also describe a straightforward numerical approach to computing the acoustic radiation force using contour integrals tailored to Cartesian grids. The method avoids computing derivatives of velocities inside or near the smeared region, where such calculations could potentially compromise the accuracy of the obtained acoustic radiation force values. For the VP approach, we solve the first-order system using MUMPS direct solver while the second-order system is solved iteratively utilizing a novel projection method-based preconditioner. For the second-order system, we study the number of iterations required to achieve a target residual for progressively large penalty factors, and show that increasing penalty factors result in a decrease in number of iterations for the preconditioned flexible GMRES (FGMRES) solver, demonstrating its efficacy. Moreover, solver scalability under grid refinement is demonstrated to show that even with large grid sizes, the number of iterations required to achieve convergence remains bounded. We test the VP method for acoustofluidic problems in both smooth and non-smooth domains containing sharp corners. By systematically comparing our results against the body-fitted solutions, we identify the optimal penalty factors and smeared region thickness for the test cases considered in this work. Lastly, we consider test cases representative of common microfluidic devices to demonstrate the efficacy of the VP technique for simulating acoustofluidic devices. The numerical test cases presented in this work will facilitate further adoption of the VP method for modeling acoustically-driven flows and will pave the way for simulating immersed bodies' motion within such flows.

\section{Continuous equations of motion}
Let $\Omega \subset \mathbb{R}^d$ represent a fixed region of space in spatial dimensions $d$ = 2  or 3.
The volume penalized compressible Navier-Stokes equations governing the interaction of acoustics with an immersed object, and the thermodynamic equation of state read as
\begin{subequations} 
\begin{alignat}{2}
    &\frac{\partial \rho}{\partial t}+\bnabla \cdot(\rho \v)=0, \label{eq:mass} \\
    &\frac{\partial (\rho \v)}{\partial t}+\bnabla \cdot(\rho \v \otimes \v)=\bnabla \cdot \bsigma + \fb, \label{eq:momentum}\\
    &p=\pfunc(\rho), \label{eq:state}
\end{alignat}
\end{subequations} 
in which $\rho$ is the spatially-varying mass density, $\v$ is the Eulerian fluid velocity, $p$ is the fluid pressure, $\bsigma$ is the Cauchy stress tensor and $\vec{f}_\textrm{b}$ is the Brinkman penalty force. The above equations hold over the entire computational domain $\Omega$. The domain $\Omega$ is further split into two non-overlapping regions: $\Omegaf (t) \subset \Omega $ occupied by the fluid, and $\Omegab (t) \subset \Omega $ occupied by the immersed body, such that $\Omega=\Omegaf (t)\cup \Omegab (t)$. We consider a linear, viscous (Newtonian) fluid such that $\bsigma$ can be expressed as
\begin{align}
    \bm{\sigma}=-p \V{I}+\mu(\nabla \v+(\nabla \v)^\intercal)+\lambda (\nabla \cdot \v) \V{I},\label{eq:sigma}
\end{align}
in which $\mu$ and $\lambda$ denote spatially-varying shear and bulk viscosity, respectively. \Revone{Following prior studies~\cite{kshetri2025consistent,nama2014investigation,muller2012numerical}, we have retained the last term in Eq.~\eqref{eq:sigma} to fully account for viscous attenuation of the acoustic wave, both within and outside the boundary layer.}
The Brinkman penalty force is written as
\begin{equation}
  \fb = \frac{\chi}{\kappa} (\vb-\v), \label{eq:penalty_force}
\end{equation}
such that it imposes the structure velocity $\vb$ onto the fictitious fluid contained within $\Omegab (t)$. Physically, this amounts to treating the immersed body as a porous region with vanishing permeability $\kappa \ll 1$. The immersed body is identified via an indicator function $\chi$ that takes the value one inside $\Omegab (t)$ and zero outside. 
Here, we consider finite but small values of $\kappa$ to model the presence of a rigid solid within an acoustically-actuated microfluidic system.

\subsection{Perturbation approach}
We consider a perturbation expansion approach to linearize the governing equations by expanding the fluid velocity, density, and pressure as an infinite series 
\begin{subequations} 
\begin{alignat}{2}
& \v=\v_0+\epsilon \v'+\epsilon^2 \v''+\mathcal{O}(\epsilon^3),\\
& p=p_0+\epsilon p'+\epsilon^2 p''+\mathcal{O}(\epsilon^3),\\
& \rho=\rho_0+\epsilon \rho'+\epsilon^2\rho''+\mathcal{O}(\epsilon^3),
\end{alignat}
\end{subequations} 
in which $\epsilon$ is a non-dimensional smallness parameter typically defined as $\epsilon = d/a$ with $d$ denoting acoustic displacement amplitude and $a$ is the characteristic length. We introduce the notation $A_1=\epsilon A'$ and $A_2=\epsilon^2 A''$  for the primary unknown fields $A$ such that the subscripts denote the order of the respective fields. The zeroth-, first-, and second-order fields denote the system's unperturbed, oscillatory, and mean response, respectively. Here,  we take the unperturbed fluid to be quiescent and set $\v_0 = 0 $.

Substituting the perturbation expansion of the primary fields into the governing equations (Eqs.~\eqref{eq:mass}--~\eqref{eq:penalty_force}), and collecting terms of $\mathcal{O}(\epsilon)$ yields the following first-order (oscillatory) system of equations 
\begin{subequations} \label{eq:first-order-eqs}
\begin{alignat}{2}
&\frac{\partial \rho_1}{\partial t}+\bnabla \cdot\left(\rho_0 \vone\right)=0,\label{eq:first-mass}\\
&\frac{\partial (\rho_0 \vone)}{\partial t}=\bnabla \cdot \bm{\sigma}_1 + \fbone, \label{eq:first-momentum}\\
&p_1 = c_0^2\rho_1, \label{eq:first-state}\\
&\bm{\sigma}_1=-p_1 \V{I}+\mu(\bnabla \vone+(\bnabla \vone)^\intercal)+\lambda (\bnabla \cdot \vone) \V{I},\label{eq:first-sigma}\\
&\fbone = \frac{\chi}{\kappa} (\voneb - \vone),\label{eq:first-penalty}
\end{alignat}
\end{subequations} 
in which $c_0$ denotes the speed of sound in the fluid. Following prior works, we take the first-order fields to be periodic in time with time period $T$ and angular frequency $\omega = 2\pi/T$. Accordingly, we seek solutions of the form 
\begin{equation}
A_1(\r, t) =  \hat{A}_1(\r)e^{\iota \omega t} = (A_1^{\rm r} (\r) + \iota A_1^{\rm i} (\r))e^{\iota \omega t},\label{eq:first-order-harmonic}
\end{equation}
in which $A_1$ denotes the first-order fields ($\pone$, $\rhoone$, or $\vone$) and the superscripts `$\rm r$' and `$\rm i$' denote the real (Re) and imaginary (Im) components of the spatially-varying part of a first-order quantity, respectively.

Repeating the same approach to gather $\mathcal{O}(\epsilon^2)$ terms, followed by a temporal averaging operation of the form $\langle A\rangle=\frac{1}{T} \int_T A\; \d t$ over the acoustic time period $T$, the second-order system of equations is obtained as follow
\begin{subequations}
\begin{alignat}{2}
&\bnabla \cdot \langle\rho_0 \vtwo\rangle=-\bnabla \cdot\langle\rho_1 \vone\rangle,\label{eq:second-mass}\\
&\bnabla \cdot (\langle\bm{\sigma}_2\rangle-\langle\rho_0\vone \otimes \vone\rangle)+ \fbtwo=\vec{0},\label{eq:second-mom}\\
&\langle\bm{\sigma}_2\rangle=-\langle p_2 \rangle \V{I}+\mu(\bnabla \langle \vtwo \rangle+(\bnabla \langle\vtwo\rangle)^\intercal)+\lambda (\nabla \cdot \langle\vtwo\rangle) \V{I},\\
&\fbtwo = \frac{\chi}{\kappa} (\vtwob - \vtwo).\label{eq:second-penalty}
\end{alignat}
\end{subequations}

We refer the reader to our recent work~\cite{kshetri2025consistent} for a detailed derivation of the first- and second-order equations. This work differs from our previous study only in that the Brinkman penalty force $\fb$ is incorporated into the momentum Eq.~\eqref{eq:momentum}. Since this term is linear in velocity, it results in a straightforward decomposition into first- and second-order components. \Revone{Therefore, the first-order system comprises Brinkman penalized coupled Hemlholtz equations, while the second-order equations represent a steady-state, low-Mach Stokes system with an additional Brinkman penalty term.}

\subsection{Boundary conditions}
The test cases presented in this work concern micro-acoustofluidic devices where the typical domain of interest comprises a liquid-filled microfluidic channel with immersed objects, if any. The system's response under the action of acoustic waves is typically modeled by prescribing a known time-periodic oscillatory motion to one or more boundaries of the domain. While the fixed boundaries of the domain are prescribed the usual no-slip boundary condition, the treatment of oscillating boundaries deserves careful attention. Specifically, the displacement of the oscillating boundaries is usually prescribed in a Lagrangian sense as being $\presdisp (\r_0,t)$, such that the position of the deformed surface of the oscillatory boundary is given as $\r=\r_0 + \presdisp (\r_0,t)$. In contrast, the fluid is typically modeled via an Eulerian approach and requires the prescription of the velocity at the mean position of the boundary. As such, the no-slip boundary condition needs to be expanded around the mean boundary position via a Taylor series expansion. We refer the reader to Bradley~\cite{bradley1996acoustic} and our recent work~\cite{kshetri2025consistent} for details of this expansion, and reproduce here the appropriate boundary conditions at the oscillating boundaries for the first- and second-order problem:
\begin{subequations}
\begin{alignat}{2}
    &\vone (\r_0,t) = \dot{\presdisp}(\r_0,t) ,\\
    & \vtwo (\r_0,t)=-\v^\textrm{SD}=- \langle\bnabla \vone (\r,t)\Bigr\rvert_{\r=\r_0}  \presdisp (\r_0,t) \rangle,\label{eq:SDTS}
\end{alignat}
\end{subequations}
where $\v^\textrm{SD}$ is referred to as the Stokes drift that links the fluid's Eulerian velocity to its Lagrangian velocity defined in the generalized Lagrangian mean theory~\cite{buhler2014waves,nama2017acoustic} as
\begin{equation}
    \v^\textrm{L}:=\vtwo + \v^\textrm{SD}.\label{eq:vL}
\end{equation}
Therefore, $\v^\textrm{SD}$ represent the second-order corrections to the fluid's Eulerian velocity that arise due to the oscillations of the boundary and Eq.~\eqref{eq:SDTS} imposes a zero Lagrangian velocity at the oscillating boundary.

The volume/Brinkman penalized rigid solid is modeled by prescribing $\voneb = 0$ within $\Omegab$, thereby enforcing the no-slip boundary condition at the rigid surface. In this work, we consider the immersed object to be stationary, but this assumption can be easily relaxed under the volume penalized framework.  For the second-order system, $\vtwob = -\v^{\textrm{SD}}$ is prescribed within the volume penalized solid region $\Omegab$ to enforce zero Lagrangian velocity at the rigid surface. 


\subsection{Acoustic radiation force}
The notion of acoustic radiation force on an immersed object is a central concept in nonlinear acoustics. Consider a time-dependent immersed domain $\Omegab(t)$ within an acoustically-actuated fluid domain $\Omega$. Let $S(t)$ denote the boundary of $\Omegab(t)$, with outward unit normal $\n$. From the Cauchy theorem, the net hydrodynamic force acting on $S(t)$ is given as
\begin{equation}
    \V{F}^\textrm{hydro} = \iint_{S(t)}\bsigma\,\n \,\d S, \label{eq:f_hydro}
\end{equation}
in which $\bsigma$ is the Cauchy stress in the fluid, and the integration is performed on the time-varying surface $S(t)$. Given the time scale separation in the system's oscillatory and mean response, the time-average of Eq.~\eqref{eq:f_hydro} over a time period $T$ is considered and is referred to as the \emph{acoustic radiation force}
\begin{equation}
    \frad = \langle \iint_{S(t)}\bsigma\,\n \,\d S\rangle.\label{eq:f_hydro_avg}
\end{equation}
Using the perturbation approach in conjunction with the conservation of linear momentum, Eq.~\eqref{eq:f_hydro_avg} can be expressed as~\cite{doinikov1994acoustic}
\begin{equation}
\label{eq:frad_final}
    \frad = \iint_{S_\infty}\left\langle(\sigmatwo-\rho\vone\otimes\vone\right)\n_\infty\rangle\,\d S, 
\end{equation}
in which the integration is performed over any stationary surface $S_\infty$ encompassing the immersed body $\Omegab(t)$. $\n_\infty$ denotes the outward unit normal to $S_\infty$. \Revtwo{While Eq.~\eqref{eq:frad_final} is commonly used in the literature to calculate the radiation force~\cite{doinikov1994acoustic}, we have included a detailed derivation  in~\ref{appendix:ARF} for reference.}

\section{Discretized equations of motion}
\subsection{Spatial discretization}
\label{sec_spatial_descritization}

\begin{figure}
    \centering
    \includegraphics[width=\linewidth]{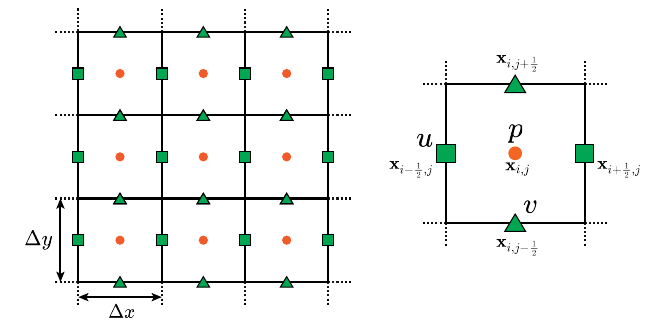}
    \caption{Schematic of a 2D staggered Cartesian grid. The left half shows the coordinate system for the staggered grid with $\Delta x$ and $\Delta y$ being the grid spacing along the $x$ and $y$ directions, respectively. The right half illustrates a single grid cell with (first- and second-order) velocity components $u$ ({\color{Green} $\blacksquare$}) and $v$
({\color{Green} $\blacktriangle$}) approximated at the cell faces  and scalar  pressure approximated at the cell center ({\color{orange} $\bullet$}).}
    \label{fig_discretized_staggered_grid}
\end{figure}

We use a uniform staggered Cartesian grid to discretize the continuous equations of motion by dividing the computational domain $\Omega$ into $N_x \times N_y$ rectangular cells. The cell size in the $x$ and $y$ directions is denoted $\dx$ and $\dy$, respectively, as illustrated in Fig.~\ref{fig_discretized_staggered_grid}.
A uniform grid spacing  $\Delta x = \Delta y = \Delta$  is used for all numerical results presented in this work. The bottom-left corner of the rectangular computational domain $\Omega$ is taken as the origin $(0, 0)$ such that the position of each center of the grid cell is given by $\x_{i,j} = \left((i + \half)\dx,(j + \half)\dy\right)$, where $i = 0, \ldots, N_x - 1$ and $j = 0, \ldots, N_y - 1$. Accordingly, the face center in the $x$-direction, which is located half a grid space away from the cell center $\x_{i,j}$ in the negative $x-$direction, is represented by $\x_{i-\half,j} = \left(i\dx,(j + \half)\dy\right)$, where $i = 0, \ldots, N_x$ and $j = 0, \ldots, N_y - 1$. Similar conventions apply to other face center locations. The $x$-component of both the first- and second-order velocities ($u_1$ and $u_2$, respectively) are stored at the centers of  $x-$direction cell faces, while the $y$-components ($v_1$ and $v_2$, respectively) are stored at the centers of the faces of the $y$-direction cell, as shown in Fig.~\ref{fig_discretized_staggered_grid}. The first- and second-order pressure values ($p_1$ and $p_2$, respectively) are stored at the cell centers.  All material properties (density $\rho$, shear viscosity $\mu$ and bulk viscosity $\lambda$) are stored at the cell centers. 
Second-order interpolation is used to interpolate the cell-centered quantities to faces and nodes, as required by the discrete spatial operators. Standard second-order finite differences are employed to approximate spatial differential operators. The spatial discretizations of the key continuous operators are as follows:
\begin{itemize}
\item The density-weighted divergence of the velocity field
$\v = (u,v)$ is approximated at cell centers by
\begin{subequations}
\begin{alignat}{2}
\label{eq_div_fd}
& \Drho(\v) = D^x_\rho \, u + D^y_\rho \, v, \\
&(D^x_\rho \, u)_{i,j} = \frac{\rho_{i+\half, j}u_{i+\half, j} - \rho_{i-\half, j} u_{i-\half, j}}{\dx}, \\
&(D^y_\rho \, v)_{i,j} = \frac{\rho_{i, j+\half} v_{i, j+\half} - \rho_{i, j-\half}v_{i, j-\half}}{\dy}. 
\end{alignat}
\end{subequations}
\item The gradient of cell-centered pressure is approximated at cell faces as
\begin{subequations}
\begin{alignat}{2}
\label{eq_grad_fd}
& \G p = (G^x p, G^y p), \\
&(G^x p)_{i-\half,j} = \frac{p_{i,j} - p_{i-1,j}}{\dx}, \\
&(G^y p)_{i,j - \half} =\frac{p_{i,j} - p_{i,j-1}}{\dy}. 
\end{alignat}
\end{subequations}
\item The continuous form of divergence of viscous strain rate tensor, which couples the velocity components through spatially variable shear viscosity, is given by
\begin{equation}
\label{eq_visc_cont}
\bnabla \cdot \left[\mu \left(\bnabla \v + (\bnabla \v)^\intercal\right) \right] = \left[
\begin{array}{c}
 (\Lmu \v)^x_{i-\half,j} \\
 (\Lmu \v)^y_{i,j-\half}  \\
\end{array}
\right] = 
\left[
\begin{array}{c}
 2 \D{}{x}\left(\mu \D{u}{x}\right) + \D{}{y}\left(\mu\D{u}{y}+\mu\D{v}{x}\right) \\
 2 \D{}{y}\left(\mu \D{v}{y}\right) + \D{}{x}\left(\mu\D{v}{x}+\mu\D{u}{y}\right) \\
\end{array}
\right].
\end{equation}
The shear viscosity operator is discretized using standard second-order, centered finite differences
\begin{subequations}
\begin{alignat}{2}
 (\Lmu \v)^x_{i-\half,j} &= \frac{2}{\dx}\left[\mu_{i,j}\frac{u_{i+\half,j} - u_{i-\half,j}}{\dx} -
					        \mu_{i-1,j}\frac{u_{i-\half,j} - u_{i-\3half,j}}{\dx}\right] \nonumber \\ 
                    &+ \frac{1}{\dy}\left[\mu_{i-\half, j+\half}\frac{u_{i-\half,j+1} - u_{i-\half,j}}{\dy} - 
					         \mu_{i-\half, j-\half}\frac{u_{i-\half,j} - u_{i-\half,j-1}}{\dy}\right] \nonumber \\
	            &+ \frac{1}{\dy}\left[\mu_{i-\half, j+\half}\frac{v_{i,j+\half} - v_{i-1,j+\half}}{\dx} - 
					         \mu_{i-\half, j-\half}\frac{v_{i,j-\half} - v_{i-1,j-\half}}{\dx}\right] \label{eq_viscx_fd} \\				         
 (\Lmu \v)^y_{i,j-\half} &= \frac{2}{\dy}\left[\mu_{i,j}\frac{v_{i,j+\half} - v_{i,j-\half}}{\dy} -
					        \mu_{i,j-1}\frac{v_{i,j-\half} - v_{i,j-\3half}}{\dy}\right] \nonumber \\ 
                    &+ \frac{1}{\dx}\left[\mu_{i+\half, j-\half}\frac{v_{i+1,j-\half} - v_{i,j-\half}}{\dx} - 
					         \mu_{i-\half, j-\half}\frac{v_{i,j-\half} - v_{i-1,j-\half}}{\dx}\right] \nonumber \\
	            &+ \frac{1}{\dx}\left[\mu_{i+\half, j-\half}\frac{u_{i+\half,j} - u_{i+\half,j-1}}{\dy} - 
					         \mu_{i-\half, j-\half}\frac{u_{i-\half,j} - u_{i-\half,j-1}}{\dy}\right] \label{eq_viscy_fd},
\end{alignat}
\end{subequations}
in which the shear viscosity is required at both cell centers and nodes of the staggered grid (i.e., $ \mu_{i\pm\half, j\pm\half}$).
Node-centered quantities are obtained via interpolation by 
arithmetically averaging the neighboring cell-centered quantities. 

\item The continuous form of divergence of bulk viscosity strain rate tensor, which couples the velocity components through spatially variable bulk viscosity, is given by
\begin{equation}
\label{eq_bulk_cont}
\div \left[\lambda (\div \v) \V{I} \right] = \left[
\begin{array}{c}
 (\Llamb \v)^x_{i-\half,j} \\
 (\Llamb \v)^y_{i,j-\half}  \\
\end{array}
\right] = 
\left[
\begin{array}{c}
 \D{}{x}\left(\lambda (\D{u}{x} + \D{v}{y}) \right)  \\
 \D{}{y}\left(\lambda (\D{u}{x} + \D{v}{y}) \right) \\
\end{array}
\right].
\end{equation}

The bulk viscosity operator is discretized using standard second-order, centered finite differences as
\begin{subequations}
\begin{alignat}{2}
 (\Llamb \v)^x_{i-\half,j} &=  \frac{1}{\dx}\left[\lambda_{i, j} \left(\frac{u_{i+\half,j} - u_{i-\half,j}}{\dx} + 
					         \frac{v_{i,j+\half} - v_{i,j-\half}}{\dy} \right) - \right. \nonumber \\ 
                             & \hspace{3em} \left. \lambda_{i-1, j} \left(\frac{u_{i-\half,j} - u_{i-\frac{3}{2},j}}{\dx} + 
					         \frac{v_{i-1,j+\half} - v_{i-1,j-\half}}{\dy} \right) \right] \label{eq_bulkx_fd} \\	         
 (\Llamb \v)^y_{i,j-\half} &= \frac{1}{\dy}\left[\lambda_{i, j} \left(\frac{u_{i+\half,j} - u_{i-\half,j}}{\dx} + 
					         \frac{v_{i,j+\half} - v_{i,j-\half}}{\dy} \right) - \right. \nonumber \\ 
                             & \hspace{3em} \left. \lambda_{i, j-1} \left(\frac{u_{i+\half,j-1} - u_{i-\half,j-1}}{\dx} + 
					         \frac{v_{i,j-\half} - v_{i,j-\frac{3}{2}}}{\dy} \right) \right] \label{eq_bulky_fd}.
\end{alignat}
\end{subequations}

\end{itemize}
\subsection{Specification of Brinkman penalized region}
A smoothed Heaviside function is used to specify the boundary of the solid region  $\partial\Omegab(t)$. The transition of the  indicator function $\chi = 1-\widetilde{H}$ occurs smoothly across thickness of the smeared region specified by $\ncells$ grid cells on either side of the interface.
The usual numerical Heaviside function is defined at the cell centers of the staggered grid:
\begin{equation}
    \widetilde{H}_{i, j} = \begin{cases} 0, & \phi_{i, j}<-\ncells h \\
    \frac{1}{2}\left(1+\frac{1}{\ncells h} \phi_{i, j}+\frac{1}{\pi} \sin \left(\frac{\pi}{\ncells h} \phi_{i, j}\right)\right), & \left|\phi_{i, j}\right| \leqslant \ncells h \\
    1, & \text{ otherwise}\end{cases}
\end{equation}
in which $h$ is a suitable measure of the cell size (e.g., $\Delta x$), and $\phi_{i,j}$ is a signed distance function whose zeroth contour denotes the fluid-solid interface. The signed distance function, by our convention, is defined to be negative inside the rigid solid region $\Omegab$ and positive outside in the fluid region $\Omegaf$. 

\subsection{Contour selection for computing acoustic radiation force}
\begin{figure}
    \centering
    \includegraphics[width=0.65\linewidth]{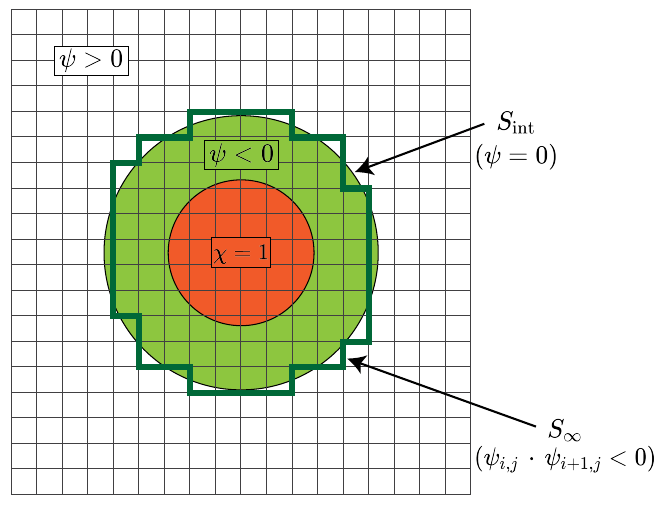}
    \caption{Schematic representation of contour selection for computing $\frad$ on the immersed object, denoted in orange. First, an arbitrary smooth contour $S_\textrm{int}$, not necessarily aligned with the Eulerian grid, is considered. Subsequently, the integration in Eq.~\eqref{eq:frad_final} is performed by representing $S_\textrm{int}$ as a stair step contour $S_\infty$ aligned with the grid cell boundaries via a level set field $\psi$.}
    \label{fig:frad_integration}
\end{figure}

The computation of $\frad$ via  Eq.~\eqref{eq:frad_final} requires the selection of an appropriate contour $S_\infty$ that encompasses the immersed object. To this end, we consider an intermediate circular contour $S_\textrm{int}$ that, in general, does not conform to the underlying Eulerian grid. In our implementation, the zeroth contour of a level set function $\psi$ denotes $S_\textrm{int}$. The integration in Eq.~\eqref{eq:frad_final} is performed by representing the circular contour in a stair step manner using the adjacent cell faces (see Fig.~\ref{fig:frad_integration}). A cell face is considered to be a part of the stair step contour if the two grid cells abutting it have the level set function $\psi$ of opposite signs. Following this identification, $\frad$ is computed by directly summing up the pressure, viscous, and momentum forces from the surrounding fluid on the areal elements of the contour $S_\infty = S_\textrm{stair}$ as
\begin{equation}
    \frad=\sum_{f \in S_\textrm{stair}} \left[ -p_2 \n_f+\mu(\nabla \v_2+(\nabla \v_2)^\intercal)\cdot \n_f+\lambda (\nabla \cdot \v_2)\n_f-\rho_0\vone(\vone\cdot \n_f) \right]\Delta A_f,
\end{equation}
in which $\n_f$ denotes the outward normal of the Cartesian face elements that comprise $S_\text{stair}$. The relevant cell-centered quantities (such as pressure $p$, viscosities $\mu$ and $\lambda$) are interpolated onto the Cartesian cell faces via arithmetic averaging. We note that any fixed, arbitrarily shaped surface that encompasses the immersed object can be chosen as the integration surface in Eq.~\eqref{eq:frad_final}. As such, a stair step representation of the contour does not constitute a numerical approximation.

\subsection{Matrix form of the equations} \label{sec_matrix}

Substituting the time periodic solution (Eqs.~\eqref{eq:first-order-harmonic}) into the first-order mass balance Eq.~\eqref{eq:first-mass} yields 
\begin{equation} \label{eq:first-mass-vec} 
 \iota \omega (\rhoone^{\rm r} + \iota \rhoone^{\rm i}) + \bnabla \cdot \left [\rho_0 (\vone^{\rm r} + \iota \vone^{\rm i}) \right] = 0.
\end{equation}
Substituting $\rhoone^{\rm r}=\pone^{\rm r}/c_0^2$ and $\rhoone^{\rm i}=\pone^{\rm i}/c_0^2$ into Eq.~\eqref{eq:first-mass-vec}, followed by separation of real and imaginary terms yields two first-order mass balance equations:
\begin{subequations} \label{eq:first-mass-comps}
\begin{alignat}{2}
    \textbf{Imaginary:}\quad \bigg(\frac{\omega}{c_0^2}\bigg) \pone^{\rm r} + \bnabla \cdot (\rho_0 \vone^{\rm i}) &= 0, \label{eq:im_cont_first}\\
    \textbf{Real:}\quad -\bigg(\frac{\omega}{c_0^2}\bigg) \pone^{\rm i} + \bnabla \cdot (\rho_0 \vone^{\rm r}) &= 0. \label{eq:re_cont_first}
\end{alignat}
\end{subequations}
Similarly, substituting the time periodic solution (Eqs.~\eqref{eq:first-order-harmonic}) into the first-order momentum Eq.~\eqref{eq:first-momentum} yields 
\begin{equation}
\begin{split}
    \iota \omega \rho_0 ( \vone^{\rm r} + \iota \vone^{\rm i} )  = - &\bnabla(\pone^{\rm r} + \iota \pone^{\rm i}) + \bnabla \cdot [\mu (\bnabla ( \vone^{\rm r} + \iota \vone^{\rm i} ) + (\bnabla( \vone^{\rm r} + \iota \vone^{\rm i} ))^\intercal) ]\\
    &+ \bnabla \left [\lambda \bnabla \cdot ( \vone^{\rm r} + \iota \vone^{\rm i} ) \right]+\frac{\chi}{\kappa}\left[(\voneb^{\rm r}+\iota \voneb^{\rm i}) - (\vone^{\rm r}+\iota \vone^{\rm i})\right].    
\end{split}
\end{equation}
Separating the real and imaginary parts gives two first-order momentum equations:
\begin{subequations} \label{eq:first-momentum-comps}
\begin{alignat}{2}
    \textbf{Imaginary:}\quad  \omega \rho_0 \vone^{\rm r} &= -\bnabla \pone^{\rm i}+\bnabla \cdot[\mu(\bnabla \vone^{\rm i}+(\bnabla \vone^{\rm i})^\intercal)]+\nabla[\lambda \bnabla \cdot \vone^{\rm i}]+\frac{\chi}{\kappa}(\voneb^{\rm i}- \vone^{\rm i}), \\
    \textbf{Real:}\quad -\omega \rho_0 \vone^{\rm i} &= -\bnabla \pone^{\rm r}+\bnabla \cdot[\mu(\nabla \vone^{\rm r}+(\bnabla \vone^{\rm r})^\intercal)]+\nabla[\lambda \bnabla \cdot \vone^{\rm r}]+\frac{\chi}{\kappa}(\voneb^{\rm r} - \vone^{\rm r}).
\end{alignat}
\end{subequations}
Eqs.~\eqref{eq:first-momentum-comps} and~\eqref{eq:first-mass-comps}, when written in a matrix form, read as
\begin{equation} \label{eq:matrix-first-brinkman}
\begin{bmatrix}
\omega \vrho_0 &  \V{\chi} / \kappa + \V{L_{\mu+\lambda}}  & \V{0} & \G\\
\V{\chi} / \kappa + \V{L_{\mu+\lambda}}  & -\omega \vrho_0 & \G & \V{0}\\
\V{0} & \Drho_0 & \frac{\omega}{c_0^2}\V{I} & \V{0}\\
\Drho_0 & \V{0} & \V{0} & -\frac{\omega}{c_0^2}\V{I}
\end{bmatrix}  
\begin{bmatrix}
\vone^{\rm r} \\
\vone^{\rm i} \\
\vpone^{\rm r} \\
\vpone^{\rm i}
\end{bmatrix} 
=
\frac{\V \chi}{\kappa} \begin{bmatrix}
\voneb^{\,\rm i}  \\
\voneb^{\,\rm r}   \\
\V{0} \\
\V{0}
\end{bmatrix} .
\end{equation}
in which $\vrho_0$ and $\V{\chi}$ denote the zeroth-order density field and indicator function, respectively, that are interpolated to the face centers. The discrete versions of the spatial operators in Eq.~\eqref{eq:matrix-first-brinkman} are defined as
\begin{subequations} 
\begin{alignat}{2}
    & \Drho_0(\v)=\bnabla \cdot (\rho_0 \v),\\
    & \V{L_{\mu+\lambda}}(\v) = -\V{L_\mu} (\v) -\V{L_\lambda} (\v) =-\bnabla \cdot [\mu (\bnabla \v+(\bnabla \v)^\intercal)+\lambda (\bnabla \cdot \v) \V{I})],\\
    & \V{G}(\V{p})= \bnabla p.
\end{alignat}
\end{subequations}
At the second-order, we have a steady-state, low-Mach, and Brinkman penalized Stokes system, which in matrix form $\V{A} \x = \b$ reads as 
\begin{align} \label{eq:matrix-second-brinkman}
\begin{bmatrix}
\V{\chi} / \kappa + \V{L_{\mu+\lambda}} & \G\\
-\Drho_0 & \V{0}
\end{bmatrix}  
\begin{bmatrix}
\vtwo \\
\vptwo
\end{bmatrix} 
&=
\begin{bmatrix}
-\langle \Drho_0(\vone \otimes \vone)\rangle \\
\Drho_0(\v^{\rm SD})
\end{bmatrix}
+
\frac{\V{\chi}}{\kappa}
\begin{bmatrix}
\vtwob \\
\V{0}
\end{bmatrix}.
\end{align}
In Eq.~\eqref{eq:matrix-second-brinkman}, $\vtwob = -\v^{\rm SD}$ is prescribed within the solid domain $\Omegab$, where $\v^{\rm SD}=\langle(\bnabla \vone)   \presdisp \rangle$ with $\presdisp=\vone/(\iota \omega)$ such that its components are given as
\begin{subequations}
\label{eq:vsd_comp}
    \begin{alignat}{2}
        v^{\textrm{SD}}_x &= \frac{1}{2 \omega}\left (\frac{\partial u_1^{r}}{\partial x}u_1^{i} + \frac{\partial u_1^{r}}{\partial y}v_1^{i} - \frac{\partial u_1^{i}}{\partial x}u_1^{r} - \frac{\partial u_1^{i}}{\partial y}v_1^{r} \right ),\\
        v^{\textrm{SD}}_y &= \frac{1}{2\omega}\left (\frac{\partial v_1^{r}}{\partial x}u_1^{i} + \frac{\partial v_1^{r}}{\partial y}v_1^{i} - \frac{\partial v_1^{i}}{\partial x} u_1^{r} - \frac{\partial v_1^{i}}{\partial y}v_1^{r} \right ). 
    \end{alignat}
\end{subequations}
Accordingly, we first obtain the velocity field $\vone$ by solving Eq.~\eqref{eq:matrix-first-brinkman}. The gradients of $\vone$ are then used to obtain the spatially-varying Stokes drift velocity via Eq.~\eqref{eq:vsd_comp} which, in turn, is used to prescribe $\vtwob$ on the right hand side of the second-order system (Eq.~\eqref{eq:matrix-second-brinkman}).

Referring to first- and second-order momentum equations (see Eqs.~\eqref{eq:matrix-first-brinkman} and \eqref{eq:matrix-second-brinkman}), the magnitude of ${\kappa}^{-1}$ should scale as $\omega \rhozero$ and as $(\mu+\lambda)/h^2$, respectively. In practice, we use a much stronger penalty to impose velocity boundary condition in both systems. In our implementation, ${\kappa}^{-1}$ is taken as $p_\kappa \omega \rhozero$ for the first-order system and as $p_\kappa (\mu+\lambda)/h^2$ for the second-order system, where $p_\kappa$ is a case-specific \emph{penalty factor}.

\subsection{Linear solvers}
The first-order system of Eqs.~\eqref{eq:matrix-first-brinkman} represents coupled Helmholtz equations in the complex amplitude of $\vone$. We solve the first-order system with MUMPS~\cite{MUMPS:1}, which is a sparse direct solver to compute $\vonehat$ and $\ponehat$. Developing an efficient iterative solver for coupled Helmholtz equations is a challenging task which, to our knowledge, has not yet been addressed in the literature. 

\begin{figure}
    \centering
    \includegraphics[width=0.35\linewidth]{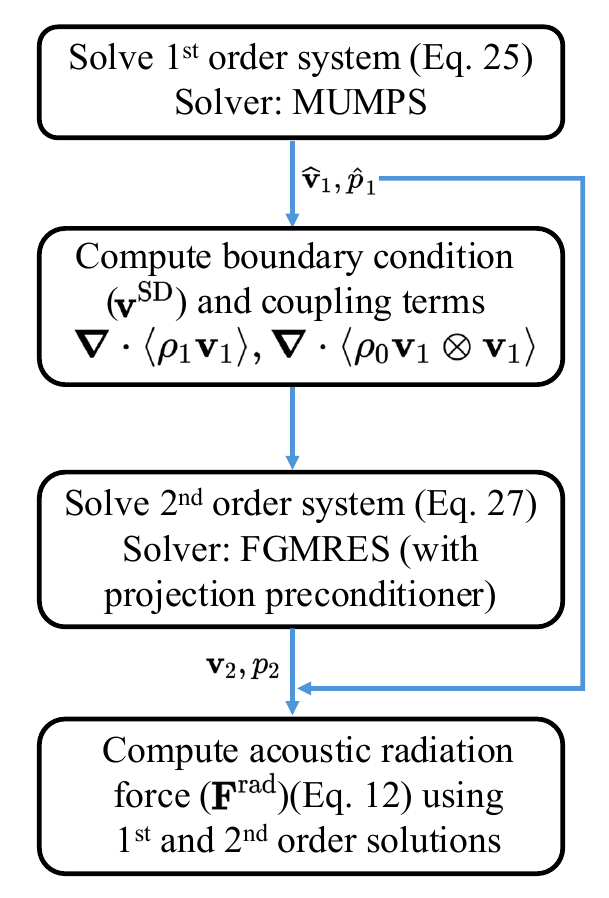}
    \caption{\Revone{Flowchart summarizing the solution algorithm.}}
    \label{fig:flowchart_solver}
\end{figure}

The second-order system of equations represent a steady-state, low-Mach Stokes system with an additional Brinkman penalty term. In our recent work we described a projection method-based preconditioner to solve the steady-state, low-Mach Stokes system. In this work we modify the projection algorithm to account for the additional Brinkman term. The  projection preconditioner is combined with the flexible GMRES (FGMRES) solver to solve Eqs.~\eqref{eq:matrix-second-brinkman} for $\vtwo$ and $p_2$. Fig.~\ref{fig:flowchart_solver} shows a flowchart summarizing our solution approach.

Traditionally, the projection method is used to solve time-dependent Navier-Stokes equations in the literature. We use the projection method as a preconditioner, not as a solver, for solving the steady-state, low Mach, and Brinkman penalized Stokes system. As a preconditioner, the projection method improves the current iterate of the outer Krylov solver by solving a residual equation of the form:
\begin{align}
\left[
\begin{array}{cc}
 \V{\chi} / \kappa + \V{L_{\mu+\lambda}} & \G\\
 -\Drho_0 & \mathbf{0} \\
\end{array}
\right]
\left[
\begin{array}{c}
  \ev\\
  \ep \\
\end{array}
\right] & =
\left[
\begin{array}{c}
 \rv\\
 \rp \\
\end{array}
\right].
\end{align}
The left-hand side vectors $\ev$ and $\ep$ are errors in the second-order velocity and pressure degrees of freedoms, respectively, and the right-hand side vectors $\rv$ and $\rp$ are residuals from the second-order momentum and mass-balance equations, respectively. 

In the first step of the projection method, an intermediate approximation to (second-order) velocity is computed by solving
\begin{equation}
\label{eq_frac_vel}
\left(\V{\chi} / \kappa + \V{L_{\mu+\lambda}} \right) \; \widetilde{\e}_{\v} = \rv.
\end{equation}
The approximation $\widetilde{\e}_{\v}$ generally does not satisfy the discrete continuity equation,
i.e., $-\Drho_0 (\widetilde{\e}_{\v}) \ne \rp$. This condition can be satisfied by introducing an auxiliary
scalar field $\vvarphi$ and carrying out an operator splitting of the form 
\begin{subequations} 
\begin{alignat}{2}
\V{\Theta}(\ev - \widetilde{\e}_{\v}) &= -\G \vvarphi, \label{eq_frac_timestep} \\
 -\Drho_0(\ev) &= \rp \label{eq_frac_continuity}.
\end{alignat}
\end{subequations}
Here, $\V{\Theta} = \V{\chi} / \kappa$ in the solid domain $\Omegab$ and $\V{\Theta} = \V{1}$ in the fluid domain $\Omegaf$. 
Taking the density-weighted divergence of Eq.~\eqref{eq_frac_timestep}, and making use of Eq.~\eqref{eq_frac_continuity} we obtain a variable-coefficient (due to $\rho_0$ and $\chi/\kappa$) Poisson equation for the scalar $\vvarphi$:
\begin{equation}
\label{eq:poisson_varphi}
-(\Drho_0 \V{\Theta}^{-1} \G) \vvarphi = -\rp - \Drho_0(\widetilde{\e}_{\v}).
\end{equation}
Using Eq.~\eqref{eq_frac_timestep}, the updated velocity solution can be computed from the solution of $\vvarphi$ as
\begin{equation}
\label{eq_frac_up_vel}
\ev = \widetilde{\e}_{\v} - \V{\Theta}^{-1} \G \vvarphi,
\end{equation}
and that of pressure can be computed as
\begin{equation}
\label{eq_frac_up_pressure}
\ep = \vvarphi.
\end{equation}
A more accurate estimate of pressure based on the spectral analysis of the Schur complement of Eq.~\eqref{eq:matrix-second-brinkman} is given by~\cite{thirumalaisamy2023pre} 
\begin{equation}
\label{eq:pressure_schur}
\ep = -( -(\Drho_0 \V{\Theta}^{-1} \G)^{-1} + 2 \V{\mu}) (\rp + \Drho_0 (\widetilde{\e}_{\v})) = \vvarphi - 2\V{\mu}(\rp + \Drho_0 (\widetilde{\e}_{\v})).
\end{equation}
Here, $\vvarphi$ and $\V{\mu}$ are cell-centered quantities. In this work we approximate pressure through Eq.~\eqref{eq:pressure_schur} within the projection preconditioner. When the Brinkman term is absent (no immersed objects in the domain), the forms of $\ev$ and $\ep$ simplifies. Those simplified expressions can be obtained by setting $\V{\Theta} \equiv \V{1}$ in Eqs.~\eqref{eq_frac_up_vel}; see also our prior work~\cite{kshetri2025consistent} for the derivation of the projection method without considering the Brinkman term.      

We set a tight relative tolerance for the outer FGMRES solver in our algorithm for solving the second-order system. A Krylov subspace is generated by the outer Krylov solver (FGMRES) by applying the action of the matrix $\A$ on vectors. It also requires the action of the projection preconditioner on residual vectors to get estimates on velocity and pressure errors. The FGMRES solver is deemed to be converged  if a value of $10^{-9}$ or below is reached for the norm of the relative residual $\mathcal{R} = \frac{||\r||}{||\b||} = \frac{||\b - \A \x||}{||\b||}$. In the projection preconditioner, we solve Eqs.~\eqref{eq_frac_vel} and~\eqref{eq:poisson_varphi} for velocity and pressure in an inexact manner. Specifically, we solve the velocity and pressure subdomain problems using a single iteration of Richardson solver that is preconditioned with a single V-cycle of a geometric multigrid solver. For both velocity and pressure problems, 3 iterations of Gauss-Seidel smoothing are performed on each multigrid level.

\subsection{Software implementation}
The volume penalized acoustofluidic solver described here is implemented within the IBAMR library~\cite{IBAMR-web-page}, an open-source C++ software enabling computational fluid dynamics algorithm development. The code is hosted on GitHub at \url{https://github.com/IBAMR/IBAMR}.
IBAMR relies on SAMRAI \cite{HornungKohn02, samrai-web-page} for Cartesian grid 
management. Solver support in IBAMR is provided by the 
PETSc library~\cite{petsc-user-ref, petsc-web-page}. 

The body-fitted grid solutions presented in the results section of this article are obtained via the commercial finite element software \comsol~6.0~\cite{comsol}. While no commercial modules exist in \comsol~for the solution of the problems formulated in this paper, the Weak PDE interface available in \comsol~allows the implementation of custom finite element problems. Specifically, we used four separate instances of the Weak PDE interface to model the first- and second-order mass and momentum equations. For both the first- and second-order problems, we used P2–P1 elements for velocity and pressure, respectively, where P1 and P2 denote triangular elements supporting Lagrange polynomials of order one and two, respectively.  Further details can be found in our prior works with body-fitted grid simulations~\cite{kshetri2025numerical,kshetri2023acoustophoresis,nama2015numerical,nama2016investigation}.


\section{Results}
In this section, we present the convergence analysis and various test cases to demonstrate the applicability and effectiveness of using the VP method for simulating acoustofluidic flows in microchannels.

\subsection{Convergence analysis}\label{sec:convg_analysis}
\begin{figure}[]
    \centering
    \includegraphics[width=0.75\linewidth]{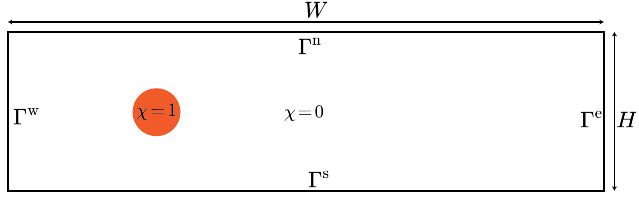}
    \caption{Schematic of a fluid-filled, rectangular microchannel comprising a rigid cylinder, denoted in orange. The channel dimensions are taken as $W = \SI{150}{\micro\metre}$ and $H= \SI{40}{\micro\metre}$. The immersed domain is circular with radius  $R_0 = \SI{10}{\micro\metre}$, and its center is located at $(W/4, H/2)$. The device is acoustically-actuated by prescribing a rectilinear actuation at the left and right walls ($\Gamma^\textrm{w} \cup \Gamma^\textrm{e}$), while the top and bottom walls ($\Gamma^\textrm{n} \cup \Gamma^\textrm{s}$) are taken to be fixed.}
    \label{fig:geom_circle}
\end{figure}

In order to test the accuracy and effectiveness of the method, we perform convergence analyses for varying mesh size ($h$), penalty factor ($p_\kappa $) and number of smeared cells ($n_\text{cells}$). Combined, these three parameters determine the accuracy in enforcing the appropriate boundary conditions at the fluid-solid interface. For all the convergence analyses presented in this section, we consider a representative case involving a rigid, solid cylinder immersed in a microfluidic channel. The computational domain is defined as a rectangle $\Omega \in [0, W] \times [0, H]$, representing the microchannel containing the circular object (see Fig.~\ref{fig:geom_circle}). The channel of width $W = \SI{150}{\micro\metre}$ and height $H = \SI{40}{\micro\metre}$ is occupied by water that is modeled as a compressible Newtonian fluid with density $\rho = 998~\mathrm{kg/m}^3$, shear viscosity $\mu = \SI{0.89}{\textrm{mPa.\second}}$, bulk viscosity $\lambda = \SI{1.88}{\textrm{mPa.\second}}$, and speed of sound $c = \SI{1500}{\meter/\second}$.  The immersed object has a radius of $R_0 = \SI{10}{\micro\metre}$, with its center located at $(x_0, y_0) = (W/4, H/2)$.
A signed distance function $\phi(\x)$ of the form 
\begin{equation}\label{eq:circle_sdf}
    \phi(\x)  = \sqrt{(x-x_0)^2 + (y-y_0)^2}- R_0.
\end{equation}
is used to define the immersed solid region/obstacle in the domain. The zeroth contour of $\phi$ denotes the fluid-solid interface. 
%

For the first-order system, the side walls ($\Gamma^\textrm{e} \cup \Gamma^\textrm{w}$) are prescribed rectilinear actuation with velocity boundary conditions $\vone = \{\iota \omega \d_0e^{\iota \omega t},0\}$, in which  $\d_0 = 1~\mathrm{nm}$ is the displacement amplitude. A frequency of $5~\textrm{MHz}$ is considered which corresponds to a half-wavelength resonance of the resulting standing wave~\cite{muller2012numerical}. The top and bottom walls ($\Gamma^\textrm{n}\cup\Gamma^\textrm{s}$) are prescribed no-slip velocity boundary conditions. Referring to Eq.~\eqref{eq:matrix-first-brinkman}, the volume penalized rigid solid is modeled by prescribing $\voneb = 0$ within $\Omegab$, thereby enforcing a zero velocity boundary condition at the rigid surface. For the second-order system, $\vtwo = -\v^{\mathrm{SD}}$ is prescribed at all channel walls to ensure zero Lagrangian velocity. This is enforced by setting $\vtwob = -\v^{\textrm{SD}}$ within the volume penalized solid region (see Eq.~\eqref{eq:matrix-second-brinkman}).

To assess the convergence of numerical solutions with respect to various parameters, we compute the acoustic radiation force ($\frad$) acting on the cylinder. Referring to Eq.~\eqref{eq:frad_final}, $\frad$ is a time-averaged, second-order quantity that depends on the primary variables and their spatial derivatives. Therefore, the convergence of $\frad$ is indicative of the convergence of the underlying first- and second-order primary variables as well as their derivatives. For each convergence parameter, we define the percentage error $\varepsilon$ in $\frad$ for a solution $g$ with respect to a reference solution $g_\textrm{ref}$ as
\begin{equation}\label{eq:errorcompute}
\varepsilon = \frac{|\frad - \frad_{\textrm{ref}}|}{|\frad_{\textrm{ref}}|}\times100,
\end{equation}
in which $\frad$ and $\frad_{\textrm{ref}}$ are the acoustic radiation force obtained from the solutions $g$ and $g_\textrm{ref}$, respectively, and $|A|$ denotes the magnitude of $A$. In all analyses presented below, the reference solution $g_\textrm{ref}$ corresponds to the solution obtained from a 
converged, body-fitted grid simulation.
\begin{figure}[]
    \centering
    \includegraphics[width=0.75\linewidth]{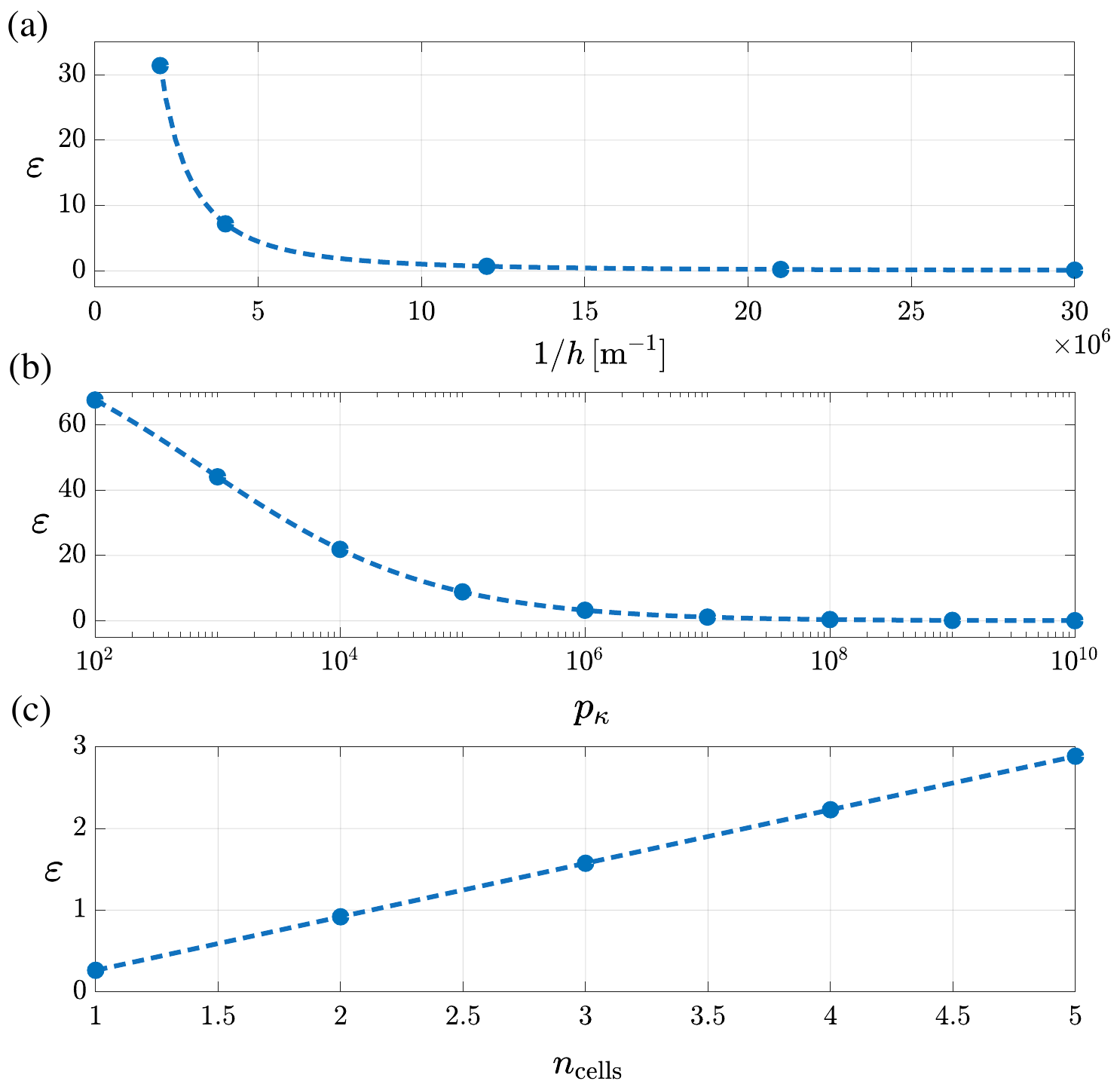}
    \caption{Convergence analysis for the acoustic radiation force $\frad$ experienced by a rigid cylinder immersed within a rectangular acoustofluidic device (see the schematic in Fig.~\ref{fig:geom_circle})  Each panel shows the variation of error in $\frad$, where the error is assessed by comparing against the body-fitted grid solution. (a) plots the error with respect to grid refinement $h$, (b) plots the error with respect to penalty factor $p_\kappa$ and (c) plots the error with respect to the number of smearing cells $\ncells$. \Revtwo{The tables containing the radiation force data for each case corresponding to Fig.~\ref{fig:convg}(a-c) are provided in~\ref{appendix:table-for-arf-data}.}}
    \label{fig:convg}
\end{figure}
%

To assess convergence with respect to grid size, we consider progressively refined grids comprising square grid cells with grid sizes $N_x\times N_y = \{300\times80, 600\times160, 1800\times480, 3150\times840,$ $4500\times1200\}$. For each grid, we evaluate the error in the acoustic radiation force by comparing the obtained value against the prediction from a body-fitted simulation. We choose $\ncells=1$ to smear the fluid-solid interface in both first- and second-order systems. Further, the penalty factor for both the first- and second-order systems is taken as $p_\kappa=10^{10}$.
Fig.~\ref{fig:convg}(a) plots the error in $\frad$ for varying grid cell sizes ($h$). As expected, the error in $\frad$ reduces with increasing mesh refinement, with an error of $0.3\%$ for the finest mesh ($h=\SI{0.033}{\micro\metre}$). Accordingly, we will use this grid refinement level to assess convergence with respect to smearing and penalty parameters in the rest of this section.
%

%


Next, we assess the errors in $\frad$ by varying the penalty factor $p_\kappa$ in the range $[10^3, 10^{10}]$. For this analysis, we take $h=\SI{0.033}{\micro\metre}$ and $\ncells=1$.  The errors in $\frad$ for different values of the penalty factor $p_\kappa$ are plotted in Fig.~\ref{fig:convg}(b). The errors show a decreasing trend with increasing penalty factors, with an error of $0.3\%$ for the largest penalty factor considered. For penalty factors greater than $10^8$, the error remains below $1\%$.
Having assessed the convergence with grid refinement and penalty factor, we now consider different levels of interfacial smearing to investigate its impact on $\frad$. For this analysis, we consider grid size $h=\SI{0.033}{\micro\metre}$ and penalty factor $p_\kappa=10^{10}$. The error decreases monotonically with the decrease in $\ncells$ (Fig.~\ref{fig:convg}(c)). This can be attributed to the fact that decreasing the smeared width of the interface leads to a more accurate representation of the interface. Overall, these results illustrate that for a suitably refined mesh, a penalty factor of $p_\kappa \geq 10^{8}$ and $ 1 \le \ncells \le 2$ are sufficient to obtain a reasonably accurate solution (within $1\%$ of the solution obtained from a body-fitted simulation).

\begin{figure}[]
    \centering
    \includegraphics[width=\linewidth]{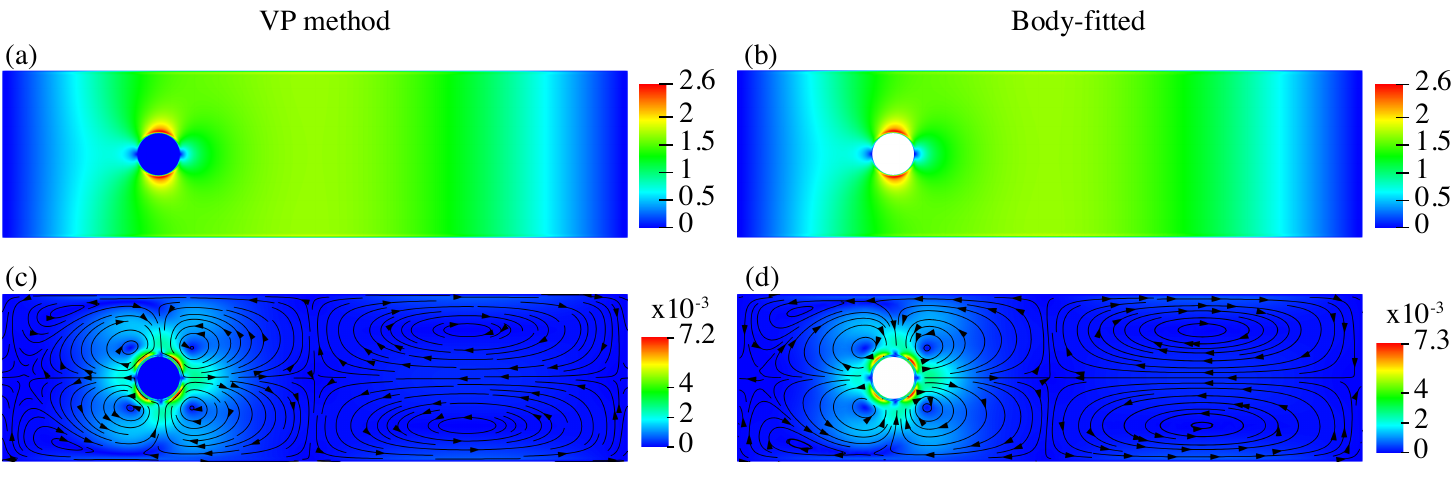}
    \caption{Comparison of the numerical solutions obtained from the VP method and body-fitted grid simulation for the rectangular acoustofluidic microchannel as illustrated in Fig.~\ref{fig:geom_circle}. (a) and (b) plot the magnitude of the first-order velocity field obtained from the VP method and body-fitted grid simulation, respectively. (c) and (d) plot the corresponding magnitudes of the second-order velocity field. The color legends in each panel indicate the velocity magnitude in m/s.} 
    \label{fig:comparison_circle}
\end{figure}

Fig.~\ref{fig:comparison_circle} compares the magnitudes of the first- and second-order velocities obtained using the VP method and the body-fitted grid simulation. The volume penalized results are obtained on a grid of size $N_x\times N_y = 4500 \times 1200$ with $p_\kappa = 10^{10}$ and $\ncells = 1$ for both first- and second-order problems. The results show excellent agreement in the first-order velocity field with the two solutions being virtually identical. For both the VP method and the body-fitted simulations, the first-order velocity exhibits maximum value in the vicinity of the immersed body, which acts as a scatterer for the acoustic field (see Figs.~\ref{fig:comparison_circle}(a) and (b)). Similarly, the solution of the steady state second-order system, which depends on the gradients of the first-order solution (see Eqs.~\eqref{eq:second-mass} and~\eqref{eq:second-mom}), also exhibits excellent agreement with a maximum error of $\approx 1\%$ in the velocity magnitude. In both Figs.~\ref{fig:comparison_circle}(c) and (d) , the streaming velocity is characterized by four symmetric vortices around the immersed object. Given the rectilinear actuation imposed in our simulations, these results are also in excellent agreement with prior results for the streaming field around a cylinder in an acoustic field~\cite{tatsuno1982secondary,kshetri2024evaluating,an2009steady,kshetri2023acoustophoresis}. Away from the scatterer, the streaming field exhibits a pair of vortices, that are characteristic of rectilinearly-actuated half-wave resonant systems, similar to those considered by Muller et al.~\cite{muller2012numerical}.

\begin{figure}
    \centering
    \includegraphics[width=0.9\linewidth]{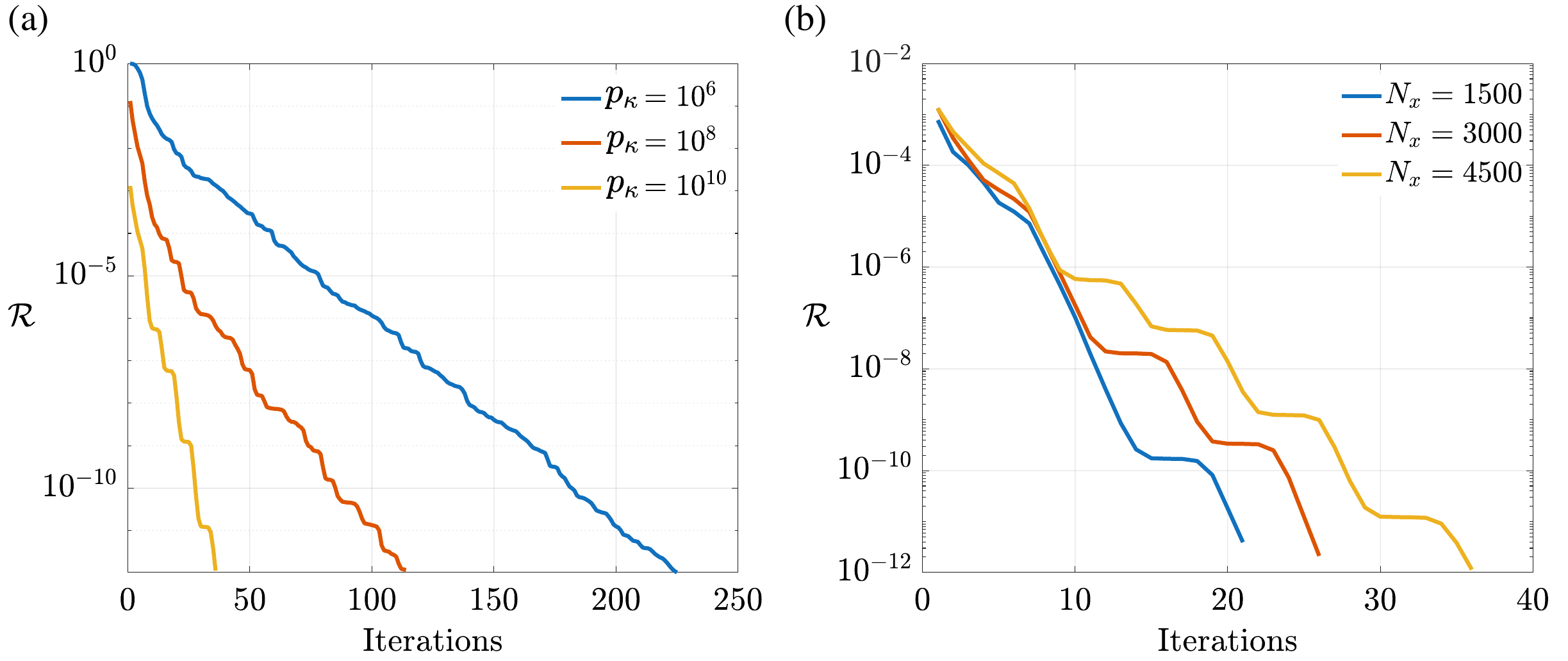}
    \caption{Convergence rate of preconditioned FGMRES solver for the second-order problem. (a) plots the error residual vs. (FGMRES) iterations for different penalty factors with a fixed grid size of $N_x\times N_y = 4500 \times 1200$. (b) plots the error residual vs. iterations for different grid sizes with a fixed penalty factor $p_\kappa = 10^{10}$.}
    \label{fig:convergence-iterations}
\end{figure}

Next, we study the performance and scalability of our preconditioner for the second-order system for varying penalty factors and grid resolution. The VP method is well known in the literature for becoming stiffer and more difficult to solve as the penalty factor increases. Fig.~\ref{fig:convergence-iterations}(a) plots the number of iterations required by our preconditioned FGMRES solver to achieve a target residual value of $\mathcal{R} = 10^{-12}$ for three different values of the penalty factor. It can be observed that as the penalty factor increases, the number of iterations to converge decreases. While this might seem surprising and contradictory to the literature, this result stems directly from the fact that an increase in the penalty factor progressively renders the velocity system diagonally dominant; see Eq.~\eqref{eq_frac_vel}. In our projection preconditioner, the Brinkman penalty is also consistently accounted for while solving the pressure Poisson equation (PPE) Eq.~\eqref{eq:poisson_varphi} via the term $\V{\Theta}$. If the Brinkman penalty is omitted from the PPE, for example by taking $\V{\Theta} \equiv \V{1}$ in Eq.~\eqref{eq:poisson_varphi}, the FGMRES solver converges very poorly. 
In practice, choosing large values of $p_\kappa$ is advantageous, as it both strengthens the enforcement of boundary conditions and accelerates convergence of the iterative solver, yielding a dual benefit. The results presented in the following two sections further demonstrate that choosing a very large penalty factor yields virtually indistinguishable results between the VP method and the body-fitted grid method, even for complex geometries featuring sharp corners.

Fig.~\ref{fig:convergence-iterations}(b) plots the convergence rate of the FGMRES solver for increasing grid resolutions with a fixed penalty factor of $p_\kappa = 10^{10}$. It can be observed that despite increasing the grid size, the number of iterations required to converge to a target residual remains bound. Combined, Figs.~\ref{fig:convergence-iterations}(a) and (b) demonstrate the efficacy and scalability of our second-order solver for large penalty factors and grid sizes. Additionally, we have included a figure in \ref{appendix:convg_additional} that shows the convergence for varying grid sizes and penalty factors in addition to the cases presented in Fig.~\ref{fig:additional_convg}(a) and (b). 

In contrast to the second-order problem, our framework solves the coupled Helmholtz first-order system with a sparse direct solver, which is not a scalable approach. Note that we solve both first- and second-order systems on distributed memory systems using the message passing interface (MPI). To our knowledge, no effective iterative solvers for the first-order problem are known in the computational acoustofluidic community and the identification of a scalable iterative solver remains a work in progress.

\subsection{Acoustic streaming in a sharp-edge microchannel}
\begin{figure}
    \centering
    \includegraphics[width=0.75\linewidth]{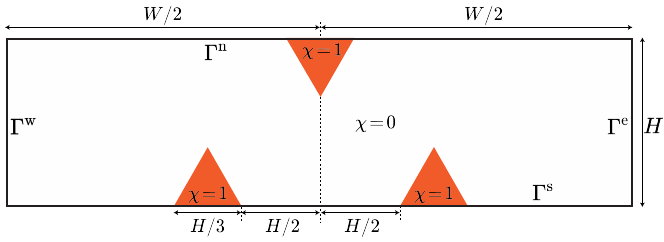}
    \caption{Schematic of a fluid-filled, sharp-edge microchannel comprising three sharp-edge structures, denoted in orange. The overall channel dimensions are taken as $W = \SI{600}{\micro\metre}$ and $H= \SI{160}{\micro\metre}$. Each sharp-edge structure is an equilateral triangle with an edge length of $H/3$. The device is acoustically-actuated by prescribing a rectilinear actuation at the left and right walls ($\Gamma^\textrm{w} \cup \Gamma^\textrm{e}$), while the top and bottom walls ($\Gamma^\textrm{n} \cup \Gamma^\textrm{s}$) are taken to be fixed.}
    \label{fig:schematic_SE}
\end{figure}

In this section, we present a comparison of the results obtained using the VP method and those obtained from a body-fitted simulation of a representative \emph{sharp-edge} acoustofluidic device. Sharp-edge acoustofluidic devices are a class of acoustofluidic devices that feature solid obstacles along the channel walls. Such devices, featuring obstructions of varying shape, have been widely employed in the acoustofluidic community for several applications, including fluid mixing, pumping, and particle patterning.~\cite{leibacher2015acoustophoretic,zhang2019acoustic,durrer2022robot,huang2013acoustofluidic,ozcelik2016acoustofluidic,pavlic2023sharp}.

\begin{figure}
    \centering
    \includegraphics[width=\linewidth]{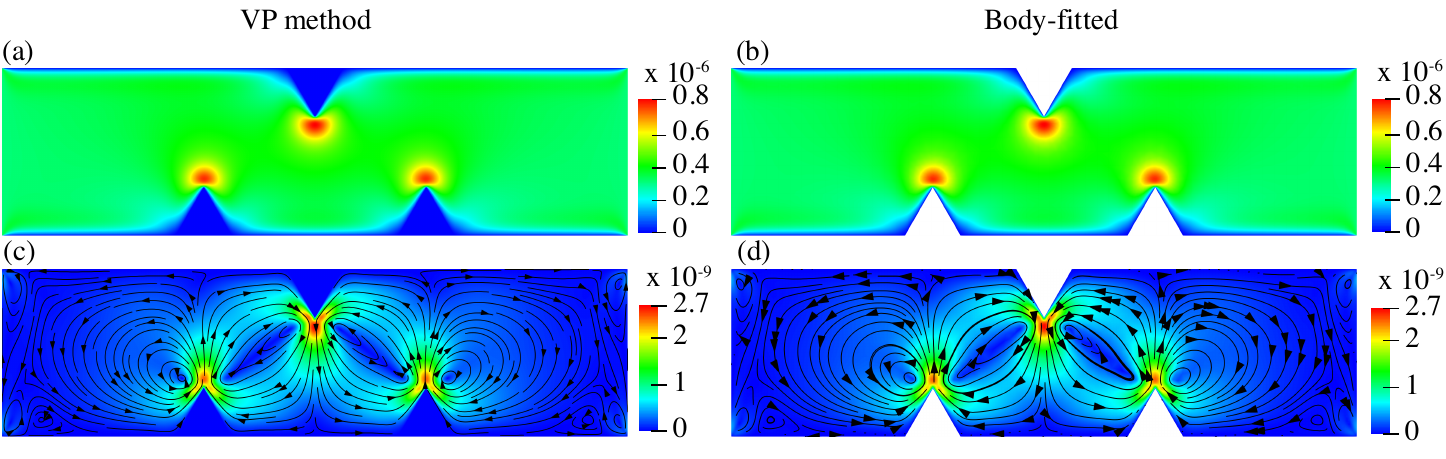}
    \caption{Comparison of the numerical solutions obtained from the VP method and body-fitted grid simulation for the sharp-edge microchannel in Fig.~\ref{fig:schematic_SE}. (a) and (b) plot the magnitude of the first-order velocity field obtained from the VP method and body-fitted grid simulation, respectively. (c) and (d) plot the corresponding magnitudes of the second-order velocity field. The color legends in each panel indicate the velocity magnitude in m/s.}
    \label{fig:SE_streaming}
\end{figure}

For this test case, we consider a microchannel geometry inspired by the sharp-edge devices reported by Huang and co-workers~\cite{huang2013acoustofluidic,huang2015acoustofluidic,huang2018sharp,nama2014investigation,junahuang2014reliable,nama2016investigation,bachman2020acoustofluidic,huang2019acoustofluidic}. Fig.~\ref{fig:schematic_SE} shows the microchannel geometry comprising three triangular sharp edges located at the top and bottom walls, with overall channel dimensions of $W = \SI{600}{\micro \meter}$ and $H = \SI{160}{\micro \meter}$. Each sharp edge domain is an equilateral triangle with an edge length of $H/3$. Referring to Fig.~\ref{fig:SE_streaming}(a), the side walls ($\Gamma^\textrm{e} \cup \Gamma^\textrm{w}$) are actuated rectilinearly with $\vone = \{\iota \omega \d_0e^{\iota \omega t},0\}$, and a frequency of 5 kHz, as is used in several experimental studies for these devices~\cite{huang2013acoustofluidic,junahuang2014reliable}. The top and bottom walls ($\Gamma^\textrm{n} \cup \Gamma^\textrm{s}$) are prescribed no-slip velocity boundary conditions. Accordingly, and similar to Sec.~\ref{sec:convg_analysis}, the volume penalized sharp-edge domains are prescribed $\voneb = 0$ and $\vtwob = -\v^{\textrm{SD}}$. A uniform Cartesian grid comprising $N_x \times N_y = 4500\times1200$ elements is considered. All other simulation parameters are the same as in Sec.~\ref{sec:convg_analysis}. 

Figs.~\ref{fig:SE_streaming}(a) and (b) compares the magnitudes of the first- and second-order velocity between the numerical solution obtained using the VP method and the body-fitted simulation. In both cases, the first-order velocity exhibits maximum values in the vicinity of the sharp edge. These results align well with previous numerical studies on similar devices and are not surprising, given that sharp edges serve as stress concentrators that amplify the velocity field under acoustic excitation. Consequently, steep gradients develop in the first-order velocity field near these edges. Since the second-order streaming flow is driven by gradients in the first-order field, this localized energy concentration leads to strong streaming velocities around the sharp edges. The streaming velocity obtained from VP method and body-fitted grid simulation are plotted in Figs.~\ref{fig:SE_streaming}(c) and (d), respectively. For both approaches, the streaming velocity is characterized by vortical flow patterns that are understood to be mixing enhancers within the acoustofluidic community. In addition, the maximum values of the first- and second-order velocity fields from the VP method and the body-fitted solution are in excellent agreement, highlighting the efficacy of our implementation in handling experimentally-relevant acoustofluidic geometries.

\Revone{Streaming velocity has been widely used as a surrogate for assessing mixing performance in acoustofluidic devices, including in our prior work on sharp-edge systems~\cite{nama2014investigation,nama2016investigation}. In these studies, mixing was evaluated by solving an advection–diffusion equation driven by the numerically computed streaming field, and experimental measurements of mixing profiles and streaming velocity obtained through tracer-particle tracking showed excellent qualitative agreement with the simulations. While the detailed flow structure depends on device geometry, actuation amplitude, and experimental setup, the flow patterns predicted by our present framework closely match experimentally reported streaming fields. Specifically, the streaming field around a single sharp edge are known to be characterized by a pair of symmetric vortices~\cite{doinikov2020acoustic,ovchinnikov2014acoustic}. For the microchannel geometry considered here, this symmetry is broken due to the presence of a sharp edge on the top wall. As such, the streaming field in Figs.~\ref{fig:SE_streaming}(c) and (d) is characterized by streamlines that traverse from the bottom sharp edge to the top sharp edge, similar to that observed experimentally by Huang et al.~\cite{huang2013acoustofluidic}.}


%


%

\subsection{Acoustic streaming in a Z-shaped microchannel}
\begin{figure}
    \centering
    \includegraphics[width=0.6\linewidth]{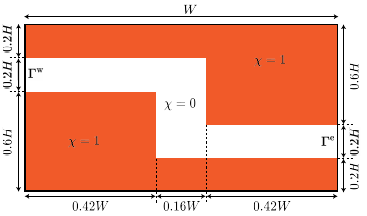}
    \caption{Schematic of a fluid-filled, Z-shaped microchannel (shown in white) carved within a larger, rectangular domain with  dimensions $W = \SI{300}{\micro \meter}$ and $H = \SI{160}{\micro \meter}$. The device is acoustically-actuated by prescribing a rectilinear actuation at the left and right boundaries of the microchannel ($\Gamma^\textrm{w} \cup \Gamma^\textrm{e}$).    
    }
    \label{fig:schematic_Lshape}
\end{figure}

\begin{figure}
    \centering
    \includegraphics[width=0.9\linewidth]{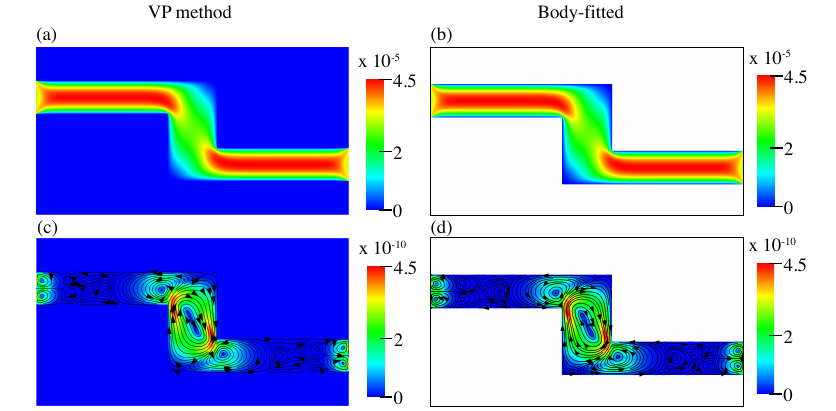}
    \caption{Comparison of the numerical solutions obtained from the VP method and body-fitted grid simulation for the Z-shaped microchannel in Fig.~\ref{fig:schematic_Lshape}. (a) and (b) plot the magnitude of the first-order velocity field obtained from the VP method and body-fitted grid simulation, respectively. (c) and (d) plot the corresponding magnitudes of the second-order velocity field. The color legends in each panel indicate the velocity magnitude in m/s. The white space outside the Z-shaped microchannel in panels (b) and (d) is only for visual purposes.}
    \label{fig:complex-channel}
\end{figure}
Next, we consider a numerical test case concerning an acoustically-actuated Z-shaped microfluidic channel embedded within a bigger Brinkman region (Fig.~\ref{fig:complex-channel}(a)). This test case is chosen to assess the performance of our implementation in scenarios where an oscillating boundary is in direct contact with a Brinkman/volume penalized region. Specifically, we want to study a scenario where a complex channel is ``carved'' within a volume penalized region. Since the second-order system requires $\vtwob=-\v^{\textrm{SD}}$ as the boundary condition, this internal flow simulation might leak fluid into the larger fictitious Brinkman region.

Referring to Fig.~\ref{fig:schematic_Lshape}, we consider a Z-shaped microchannel carved with a larger, rectangular domain with dimensions $W = \SI{300}{\micro \meter}$ and $H = \SI{160}{\micro \meter}$. The left and right boundaries of the Z-shaped microchannel ($\Gamma^\textrm{e} \cup \Gamma^\textrm{w}$) are actuated rectilinearly with $\vone = \{\iota \omega \d_0e^{\iota \omega t},0\}$, and a frequency of 5 kHz. All other boundaries of the microchannel are in contact with the volume penalized region that is prescribed $\voneb = 0$ and $\vtwob = -\v^{\textrm{SD}}$ to enforce a no-slip boundary condition at the fluid-solid interface. We consider a uniform Cartesian grid comprising $N_x \times N_y = 2250 \times 1200$ grid cells with a penalty factor $p_\kappa = 10^{10}$ and $\ncells =1$.

Figs.~\ref{fig:complex-channel}(a) and (b) show excellent agreement between the volume penalized and body-fitted grid solution for the first-order velocity. The corresponding time-averaged streaming fields are plotted in Figs.~\ref{fig:complex-channel}(c) and (d). Similar to the sharp-edge case considered in the last section, streaming vortices are observed around the sharp geometrical features ($90^\circ$ bends). A close inspection of the streaming fields in Fig.~\ref{fig:complex-channel}(c) shows that the streamlines are parallel to the internal channel walls and do not cross into the volume penalized regions. This behavior is consistent with the body-fitted grid simulations (Fig.~\ref{fig:complex-channel}(d)), where only the fluid domain is modeled and the no-slip boundary condition is imposed directly as an essential boundary condition. 
The maximum value of both the first- and second-order velocity fields obtained from the VP method agrees well with the body-fitted solution. These results illustrate the efficacy of the VP approach in simulating internal flows in geometrically complex regions.


%
%

\section{Conclusions and Discussion}
In this study, we presented a volume penalized, perturbation-based acoustofluidic solver with a second-order accurate finite difference/volume implementation. We introduce a penalty force within the compressible Navier-Stokes equations to model an acoustically-actuated fluid domain containing immersed, rigid objects. Using a perturbation approach, the nonlinear compressible Navier-Stokes equations are  segregated into a harmonic first-order problem and a time-averaged second-order problem, where the latter utilizes forcing terms and boundary conditions arising from the solution of the first-order problem. Given the linear nature of the penalty forcing, the perturbation approach results in a corresponding penalty forcing for both the first- and second-order problems to account for the presence of immersed bodies. The no-slip boundary condition at the fluid-solid interface is enforced by prescribing a known structure velocity in each order. Specifically, we prescribe a zero structure velocity for the first-order problem, while the structure velocity for the second-order problem is prescribed to be the negative of the Stokes drift, which depends on the gradient of the first-order solution. The acoustic radiation force on the immersed object is computed via integration over a stair step contour that completely encloses the immersed object.

To assess the numerical accuracy of our volume penalized finite difference/volume implementation, we compared our results against those obtained from finite element-based body-fitted grid simulations. Through a series of test cases, we demonstrate excellent agreement with the body-fitted simulation results for the primary first- and second-order fields as well as for the acoustic radiation force that depends on the gradients of these fields. We also identify suitable penalty factors and interfacial smearing widths that are sufficient to accurately capture first- and second-order solutions. Based on the test cases considered here, for a sufficiently refined mesh, a penalty factor of $p_\kappa \geq 10^{8}$ and $ 1 \le \ncells \le 2$ are adequate to obtain reasonably accurate first- and second-order solutions for acoustic streaming flows using the VP method. We also described an effective and scalable preconditioner for solving the steady state Brinkman penalized and low Mach Stokes systems iteratively in this work. Overall, these results demonstrate that our volume penalized framework can accurately capture gradients near immersed regions and highlight the efficacy of our approach in simulating acoustofluidic problems. 

\Revtwo{This work represents a key step in our ongoing efforts to develop a perturbation-based, variable-coefficient solver capable of simulating multiphase/multicomponent acoustofluidic systems. 
Our findings show that integration of a perturbation framework with the VP method, along with the appropriate splitting of boundary conditions for the first- and second-order systems, is feasible and yields excellent agreement with body-fitted grid results. This is particularly noteworthy because the second-order boundary condition depends on gradients of the first-order velocity field, which can be sensitive to errors inherent in diffuse-interface formulations such as the VP method. However, the present study focuses exclusively on stationary problems, which do not require re-gridding operations. Consequently, the advantages of employing a VP approach are not fully realized in this context. Our ongoing work focuses on investigating moving particles where IB methods are known to offer substantial benefits over body-fitted grid approaches that require frequent re-gridding. These results will be presented in a forthcoming article.
}

\Revone{
Another important direction for future research concerns the extension of the current framework to three-dimensional problems. Although our IB formulation is dimension-independent ($d=2,3$), the present implementation is constrained by the lack of scalability of the sparse direct solver used for the first-order problem. To our knowledge, scalable solvers for the coupled Helmholtz system are currently unavailable. Given the growing need to simulate experimentally observed 3D trajectories of acoustically driven microswimmers~\cite{dillinger2024steerable,ren20193d}, developing scalable iterative solvers for the coupled Helmholtz system remains a key challenge for future work.}

\section*{Acknowledgements}
This work was supported, in part, by the National Science Foundation (OAC-1931368, CBET-2234387, OIA-2229636, CBET-2407937, CBET-2407938) and the American Heart Association (23CDA1048343).

\section*{Appendix}
\begin{appendix}
\section{Evaluating the acoustic radiation force via a contour integral}
\label{appendix:ARF}

\begin{figure}
    \centering
    \includegraphics[width=0.5\linewidth]{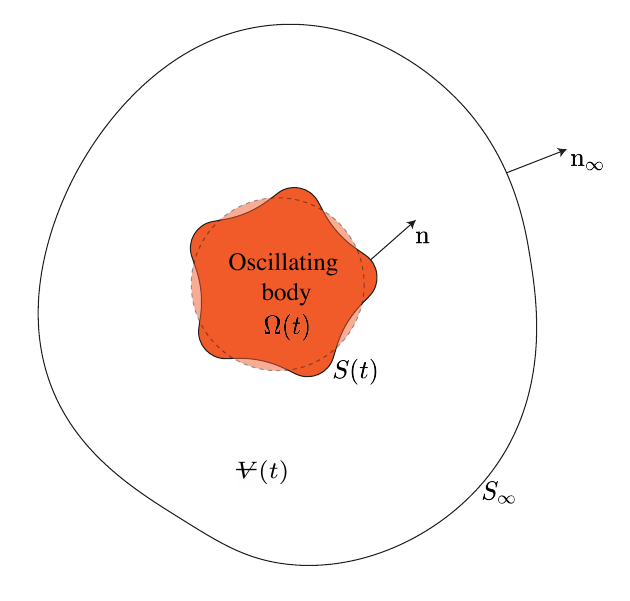}
    \caption{Schematic illustrating the computation of acoustic radiation force over an oscillating body $\Omega(t)$ with boundary $S(t)$ and outwatd unit normal $\n$. $S_\infty$ denotes an arbitrary contour, with outward unit normal $\n_\infty$, that completely encompasses the oscilalting body while $\volume(t)$ denotes the volume between $S(t)$ and $S_\infty$.}
    \label{fig:frad_derivation}
\end{figure}

Based on Eq.~\eqref{eq:f_hydro_avg}, the acoustic radiation force on a particle is defined as the time-average of the integration of Cauchy stress on the oscillating surface $S(t)$ 
\begin{equation}
\label{eq:frad_def}
    \frad = \langle\iint_{S(t)}\bsigma\,\n \,\d S\rangle.
\end{equation}
Referring to Fig.~\ref{fig:frad_derivation}, consider an arbitrary domain completely enclosing an oscillating boundary, with $\volume(t)$ denoting the volume between $S(t)$ and $S_\infty$. The balance of linear momentum over $\volume(t)$ is given as
\begin{equation}
    \frac{\partial(\rho\v)}{\partial t}+\bnabla\cdot(\rho\v\otimes\v-\bsigma)=0,
\end{equation}
which upon integration over $\volume(t)$ yields
\begin{equation}
\iiint_{\volume(t)}\left[\frac{\partial(\rho\v)}{\partial t}+\bnabla\cdot(\rho\v\otimes\v-\bsigma)\right]\, \d V=0. \label{eq:int_momentum}
\end{equation}
Using divergence theorem, Eq.~\eqref{eq:int_momentum} yields
\begin{equation}
\label{eq:frad_div}
    \iiint_{\volume(t)}\frac{\partial(\rho\v)}{\partial t} \d V+\iint_{S(t)}(\rho\v\otimes\v-\bsigma)(-\n)\, \d S+\iint_{S_\infty}(\rho\v\otimes\v-\bsigma)\n_\infty \, \d S =  0,
\end{equation}
in which the normal in the second integral is taken as $-\n$ since we require the normal to  point outward from the fluid, and $\n_\infty$ denotes the outward normal to the surface $S_\infty$. Now, consider the momentum $\rho\v$ in \st{$V$}($t$). Using the Leibniz rule, we can write
\begin{equation}
    \frac{\d}{\d t}\iiint_{\volume(t)} \rho \v \, \d V = \iiint_{\volume(t)}\frac{\partial(\rho\v)}{\partial t}\, \d V+\iint_{S(t)}\rho\v\left(\v_{S(t)}\cdot (-\n)\right) \d S+\iint_{S_\infty}\rho\v\left(\v_{S_\infty}\cdot \n_\infty\right)\, \d S,
    \label{eq:rtt}
\end{equation}
in which $\v_{S(t)}$ and $\v_{S_\infty}$ denote the velocities of the surfaces $S(t)$ and $S_\infty$, respectively. Consider $S_\infty$ to be moving with velocity $\U$ (i.e., $\v_{S_\infty} = \U$), while the particle surface $S(t)$ moves with the material velocity $\v$ (i.e., $\v_{S(t)} = \v$). Eq.~\eqref{eq:rtt} yields
\begin{equation}
    \frac{\d}{\d t}\iiint_{\volume(t)}\rho\v \, \d V = \iiint_{\volume(t)}\frac{\partial (\rho \v)}{\partial t} -\iint_{S(t)}\rho\v\left(\v\cdot\vec{n}\right) \d S +\iint_{S_\infty}\rho\v\left(\U\cdot\n_\infty\right) \d S.\label{eq:rtt_2}
\end{equation}
Subtracting Eq.~\eqref{eq:rtt_2} from Eq.~\eqref{eq:frad_div} yields
\begin{equation}
\begin{aligned}
    \bigg\{&\cancel{\iiint_{\volume(t)}\frac{\partial(\rho\v)}{\partial t}}\, \d V
    -\iint_{S(t)}\cancel{(\rho\v\otimes\v}-\bsigma)\n\, \d S
    +\iint_{S_\infty}(\rho\v\otimes\v-\bsigma)\n_\infty\, \d S\bigg\} \\
    -\bigg\{&\cancel{\iiint_{\volume(t)}\frac{\partial (\rho \v)}{\partial t}} \, \d V -\cancel{\iint_{S(t)}\rho\v\left(\v\cdot\n\right)} \, \d S
+\iint_{S_\infty}\rho\v\left(\U\cdot\n_\infty\right)\, \d S-\frac{\d}{\d t}\iiint_{\volume(t)}\rho\v \, \d V \bigg\}=0,
\end{aligned}
\end{equation}
which can be re-arranged as
\begin{equation}
        \iint_{S(t)}\bsigma \n \, \d S = \iint_{S_\infty}\left(\bsigma-\rho\v\otimes\v\right)\n_{\infty}\, \d S +\iint_{S_\infty}\rho\v\left(\U\cdot\n_{\infty}\right) \, \d S
        -\frac{\d}{\d t}\iiint_{\volume(t)}\rho\v \, \d V.
\end{equation}
Taking the time-average of the equation above, using the definition of $\frad$ (Eq.~\eqref{eq:frad_def}), and noting that the time average of the last term on the right hand side vanishes~\cite{baudoin2020acoustic}, we obtain
\begin{equation}
        \frad = \langle\iint_{S_\infty}\left(\bsigma-\rho\v\otimes\v\right)\n_{\infty} \, \d S\rangle
        +\langle\iint_{S_\infty}\rho\v\left(\U\cdot\n_{\infty}\right)\, \d S \rangle.
\end{equation}
Lastly, if the integration surface $S_\infty$ is taken to be fixed (i.e., $\U=\bm{0}$), we obtain  
\begin{equation}
        \frad = \langle\iint_{S_\infty}\left(\bsigma-\rho\v\otimes\v\right)\n_{\infty} \, \d S\rangle,
\end{equation}
which can be expressed, up to second-order accuracy, as
\begin{equation}
        \frad = \iint_{S_\infty}\langle\left(\bsigma_2-\rhozero\vone\otimes\vone\right)\n_{\infty}\rangle \, \d S.
\end{equation}
Due to the assumption that $S_\infty$ is stationary, the time-averaging operator can be moved inside the integral.  

\noindent Dropping the subscript from $\n_\infty$, the component form of $\frad$ 
\begin{equation}
    \frad = \iint_{S_\infty} \Bigl\{ \left(- p_2  \n + \mu[\bnabla \vtwo +(\bnabla \vtwo )^{\textrm{T}}]\right)\n +\lambda  (\bnabla\cdot \vtwo)\n - \langle\rho_0 \vone(\vone\cdot\n)\rangle \Bigr\} \, \d S,
\end{equation}
in two spatial dimensions reads as
\begin{subequations}
    \begin{alignat}{2}
    F_x^{\mathrm{rad}} = \iint_{S_\infty} \Bigl\{ -p_2 n_x + \mu\left(\left[\frac{\partial v_2}{\partial x}+\frac{\partial u_2}{\partial y}\right]n_y + 2\frac{\partial u_2}{\partial x}n_x\right) + \lambda \left(\frac{\partial u_2}{\partial x}+ \frac{\partial v_2}{\partial y}\right)n_x - \rho_0\langle(\vone \cdot \n)\vone\rangle_x \Bigr\} \, \d S, \\
    F_y^{\mathrm{rad}} = \iint_{S_\infty} \Bigl\{ -p_2 n_y + \mu\left(\left[\frac{\partial v_2}{\partial x}+\frac{\partial u_2}{\partial y}\right]n_x + 2\frac{\partial v_2}{\partial y}n_y\right) + \lambda \left(\frac{\partial u_2}{\partial x}+ \frac{\partial v_2}{\partial y}\right)n_y - \rho_0\langle(\vone \cdot \n)\vone\rangle_y \Bigr\} \, \d S,
    \end{alignat}
\end{subequations}
in which
\begin{subequations}
    \begin{alignat}{2}
    \langle(\vone \cdot \n)\vone\rangle_x = \frac{1}{2}\left(\left(u_1^{r}n_x+ v_1^r n_y\right)u_1^r + \left(u_1^{i}n_x+v_1^i n_y\right)u_1^i\right),\\
    \langle(\vone \cdot \n)\vone\rangle_y = \frac{1}{2}\left(\left(u_1^{r}n_x+ v_1^r n_y\right)v_1^r + \left(u_1^{i}n_x+v_1^i n_y\right)v_1^i\right).
    \end{alignat}
\end{subequations}
\section{Acoustic radiation force data corresponding to Figure~\ref{fig:convg}}
\Revtwo{
\label{appendix:table-for-arf-data}
Tables~\ref{table:arf_data},\ref{table:arf_data_b}, and \ref{table:arf_data_c} present the data corresponding to Figure~\ref{fig:convg}(a), (b) and (c) respectively. Only the x-component of the radiation force is presented, since the immersed body is positioned symmetrically along the y-direction of the channel, resulting in a zero y-component of the force.

\begin{table}[!h]
\caption{Acoustic radiation force ($\F^{\mathrm{rad}}$) data corresponding to Fig.~\ref{fig:convg}(a). $\F^{\mathrm{rad}}$ is computed for varying grid refinement ($1/h$). Error ($\varepsilon$) is computed using Eq.~\ref{eq:errorcompute}.}
\label{table:arf_data}
\centering
\begin{tabular}{ |c|c|c|c|c| }
    \hline
    grid size($h$) & $1/h$ & $\F^{\mathrm{rad}}_{\mathrm{ref}}$(N) & $\F^{\mathrm{rad}}$(N) & $\varepsilon$\\
    \hline
    $300\times80$ & 2 & 0.004 & 0.00524 & 31 \\
    $600\times160$ & 4 & 0.004 & 0.00428 & 7.5 \\
    $1800\times480$ & 12 & 0.004 & 0.00396 & 0.5 \\
    $3150\times840$ & 21 & 0.004 & 0.00398 & 0.2 \\
    $4500\times1200$ & 30 & 0.004 & 0.00399 & 0 \\
    \hline
\end{tabular}
\end{table}

\begin{table}[h!]
\caption{Acoustic radiation force ($\F^{\mathrm{rad}}$) data corresponding to Fig.~\ref{fig:convg}(b). $\F^{\mathrm{rad}}$ is computed for varying penalty factor $(p_k)$. Error ($\varepsilon$) is computed using Eq.~\ref{eq:errorcompute}.}
\label{table:arf_data_b}
\centering
\begin{tabular}{ |c|c|c|c| }
    \hline
    Penalty factor $(p_k)$ & $\F^{\mathrm{rad}}_{\mathrm{ref}}$(N) & $\F^{\mathrm{rad}}$(N) & $\varepsilon$\\
    \hline
    $10^2$ & 0.004 & 0.00129 & 67.75 \\
    $10^3$ & 0.004 & 0.00224 & 44.00 \\
    $10^4$ & 0.004 & 0.00311 & 22.25 \\
    $10^5$ & 0.004 & 0.00364 & 9.00 \\
    $10^6$ & 0.004 & 0.00388 & 3.00 \\
    $10^7$ & 0.004 & 0.00397 & 0.75 \\
    $10^8$ & 0.004 & 0.00397 & 0.75 \\
    $10^9$ & 0.004 & 0.00398 & 0.50 \\
    $10^{10}$ & 0.004 & 0.00399 & 0.25 \\
    \hline
\end{tabular}
\end{table}

\begin{table}[h!]
\caption{Acoustic radiation force ($\F^{\mathrm{rad}}$) data corresponding to Fig.~\ref{fig:convg}(c). $\F^{\mathrm{rad}}$ is computed for varying number of smearing cells ($n_{\mathrm{cells}}$). Error ($\varepsilon$) is computed using Eq.~\ref{eq:errorcompute}.}
\label{table:arf_data_c}
\centering
\begin{tabular}{ |c|c|c|c| }
    \hline
    $n_{\mathrm{cells}}$ & $\F^{\mathrm{rad}}_{\mathrm{ref}}$(N) & $\F^{\mathrm{rad}}$(N) & $\varepsilon$\\
    \hline
    1&0.004 & 0.00398 & 0.50 \\
    2 & 0.004 & 0.00396 & 1.00 \\
    3 & 0.004 & 0.00393 & 1.75 \\
    4 & 0.004 & 0.00391 & 2.25 \\
    5 & 0.004 & 0.00388 & 3.00 \\
    \hline
\end{tabular}
\end{table}
}

\Revtwo{
\section{Convergence rate of FGMRES solver}
\label{appendix:convg_additional}
To further evaluate the performance and scalability of our preconditioned FGMRES solver, we present additional results in Figure~\ref{fig:additional_convg}, which include extended convergence-rate cases beyond those shown in Figure~\ref{fig:convergence-iterations}. The observed trends further confirm that the rate of convergence improves as the penalty factor ($p_\kappa$) increases, consistent with the behavior reported in Figure~\ref{fig:convergence-iterations}. Moreover, the convergence rate remains bounded with increasing grid refinement.
\begin{figure}[h!]
    \centering
    \includegraphics[width=0.65\linewidth]{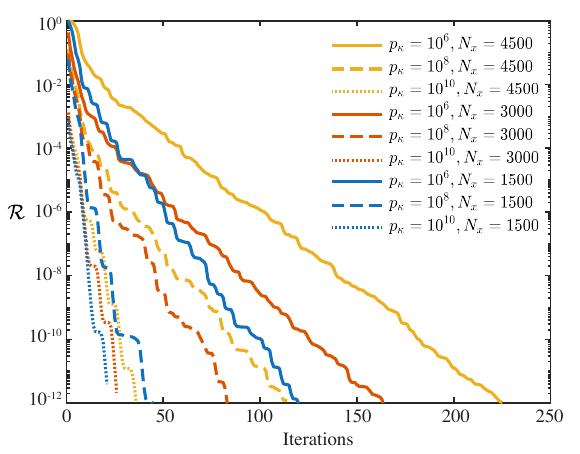}
    \caption{The plot shows the complete set of convergence-rate cases for the preconditioned FGMRES solver for the second-order problem, extending the results presented in Figure~\ref{fig:convergence-iterations}. }
    \label{fig:additional_convg}
\end{figure}
}

\end{appendix}










\bibliography{acoustic_bibliography.bib}
\end{document}